\documentclass[11pt]{article}

\usepackage{arxiv_macros}
\usepackage{tikz_macros}
\usepackage[margin=1.2in]{geometry}

\begin{document}
\raggedbottom

\title{Data-Driven Optimization of Directed Information over Discrete Alphabets}

\author{Dor Tsur\thanks{Ben Gurion University of the Negev}, Ziv Aharoni\footnotemark[1], Ziv Goldfeld\thanks{Cornell University}, and Haim Permuter\footnotemark[1]}


\date{}
\maketitle

\begin{abstract}
Directed information (DI) is a fundamental measure for the study and analysis of sequential stochastic models. In particular, when optimized over input distributions it characterizes the capacity of general communication channels. However, analytic computation of DI is typically intractable and existing optimization techniques over discrete input alphabets require knowledge of the channel model, which renders them inapplicable when only samples are available. To overcome these limitations, we propose a novel estimation-optimization framework for DI over discrete input spaces. We formulate DI optimization as a Markov decision process and leverage reinforcement learning techniques to optimize a deep generative model of the input process probability mass function (PMF). Combining this optimizer with the recently developed DI neural estimator, we obtain an end-to-end estimation-optimization algorithm which is applied to estimating the (feedforward and feedback) capacity of various discrete channels with memory. Furthermore, we demonstrate how to use the optimized PMF model to (i) obtain theoretical bounds on the feedback capacity of unifilar finite-state channels; and (ii) perform probabilistic shaping of constellations in the peak power-constrained additive white Gaussian noise channel.
\end{abstract}




\section{Introduction}
Originally proposed for the study of channels with feedback, directed information (DI) quantifies statistical and temporal dependencies between stochastic processes \cite{massey1990causality}.
It has seen a variety of applications in communications \cite{kramer1998directed}, portfolio theory \cite{permuter2008directed}, computational biology \cite{rao2006inference}, neuroscience \cite{wibral2014directed} and machine learning \cite{zhou2016causal,tiomkin2017unified}. Often, one wishes to optimize the DI over the distribution of some of the involved stochastic processes; e.g., channel capacity with and without feedback is given by the maximized DI over an appropriate set of input distributions~\cite{kim2008coding}.
However, the resulting optimization problem is usually intractable, with analytic solutions available only for a limited class of channels \cite{kim2006feedback,permuter2008capacity, sabag2015feedback,peled2019feedback}.

In the absence of analytic solutions, the capacity may be computed via DI optimization routines, such as Blahut-Arimoto-type algorithms \cite{naiss2012extension,charalambous2016directed} or methods based on Markov decision process (MDP) formulations and dynamic programming \cite{permuter2008capacity, elishco2014capacity} or reinforcement learning \cite{aharoni2020reinforcement} techniques. However, these approaches are only feasible under restrictive structural assumptions on the channel that enable tensorizing the multi-letter DI objective, e.g., unifilar\footnote{A finite-state channel is unifilar if its state evolves as a time-invariant deterministic function of the past input, output, and state tuple.} finite-state channels (FSCs) with feedback and symmetric channels. For more general channels\footnote{or when the feedback capacity itself is not algorithmically (Borel-Turing) computable \cite{grigorescu2022capacity}.}, the capacity can be bounded using the machinery of $Q$-graphs \cite{sabag2020graph, huleihel2021computable}, but tight bounds require an exhaustive search over an exponentially large space. Furthermore, all the aforementioned approaches require full knowledge of the channel probabilistic model, which is often unavailable in practice.

Empirical DI estimators may be employed to obtain approximate solutions, when the channel model is unknown but can be sampled from. For memoryless channels, joint estimation-optimization methods over continuous input spaces were proposed in \cite{letizia2021capacity, mirkarimi2021neural}. The case of channels with memory was recently treated in \cite{tsur2022neural} using the DI neural estimator (DINE) developed therein. The DINE parametrizes the Donsker-Varadhan representation of DI by recurrent neural networks (RNNs), approximates expectations by sample means, and optimizes the resulting objective over the parameter space. To compute the feedback capacity, \cite{tsur2022neural} further proposed an RNN-based generative model for continuous input distributions and jointly optimized it with DINE by propagating gradients through both models. These methods hinge on the end-to-end differentiability of the joint model, which fails to hold for discrete input alphabets. To the best of our knowledge, to date there is no known data-driven approach for optimizing (estimated) DI over discrete alphabets. The goal of this paper is to close this gap.

We propose a new method for optimizing DINE over discrete input alphabets. The input distribution is modeled by an RNN-based probability mass function (PMF) generator. We formulate the DI maximization as an MDP whose policy is modeled by the PMF generator.
Such a formulation allows us to utilize reinforcement learning techniques, and in particular, policy optimization via the policy gradients theorem \cite{sutton1999policy}. Combined with a DINE-based approximation of the MDP reward and a Monte-Carlo (MC) estimate of the policy gradient expression, this result in a tractable policy optimization objective. We then alternate between optimizing this objective and the DINE parameters, which yields an estimation-optimization procedure for the DI rate over discrete-input channels. Importantly, our approach does not rely on any knowledge of the channel transition kernel, but rather on the ability to sample its output.

We apply our approach to three main tasks concerning communication over noisy channels. First, we use the proposed method to estimate the capacity of several channels with memory. In all considered examples, the method either achieves the theoretical capacity value or converges between known upper and lower bounds. Furthermore, we show that the optimized PMF generator corresponds to known capacity-achieving input distributions. Second, we employ the generator to estimate a $Q$-graph \cite{sabag2016single}, that can be plugged into the algorithm from \cite{sabag2020graph} to obtain tight bounds on the feedback capacity of unifilar FSCs. 
Finally, we leverage the developed method to perform probabilistic shaping of pulse amplitude modulation (PAM) and quadrature amplitude modulation (QAM) constellations over peak-power-constrained additive white Gaussian noise (AWGN) channels.
Our method yields nontrivial distributions, whose information transmission rate is higher than the one obtained from a uniform distribution, which is typically used in practice \cite{proakis1994communication}.

The remainder of the text is organised as follows.
Section \ref{sec:prel} provides preliminaries and technical background.
Section \ref{sec:di_opt} derives the DI optimizer, while Section \ref{sec:implementation} discusses implementation and algorithmic aspects.
In Section \ref{sec:exp_cap} we present empirical results for channel capacity estimation.
Applications to bounding techniques for the feedback capacity of unifilar FSCs and to probabilistic shaping are the focus of Sections \ref{sec:q_graph} and \ref{sec:prob_shape}, respectively. Proofs are provided in Section \ref{sec:proofs}. Section \ref{sec:conclusion} leaves concluding remarks and discusses future directions.

\section{Background and Preliminaries}\label{sec:prel}

\subsection{Notation}
Sets are denoted by calligraphic letters, e.g., $\cX$.
When $\cX$ is finite we use $|\cX|$ for its cardinality.
For any $n\in\NN$, $\cX^n$ is the $n$-fold Cartesian product of $\cX$, while $x^n=(x_1,\dots,x_n)$ denotes an element of $\cX^n$.
For $i,j\in\ZZ$ with $i\leq j$, we use the shorthand $x_i^j:=(x_i,\dots,x_j)$; the subscript is omitted when $i=1$.
We denote by $(\Omega,\cF,\PP)$ the underlying probability space on which all random variables are defined, which is assumed to be sufficiently rich. Expectations are denoted by $\EE$; we sometime write $\EE_P$ to stress that the underlying distribution is $P$.
The set of all Borel probability measures on $\cX$ is denoted by $\cP(\cX)$ and the $k$-dimensional probability simplex is denoted $\Delta_k$.
When $\cX$ is countable, we use $p$ for the PMF associated with $P\in\cP(\cX)$. Random variables are denoted by upper-case letters, e.g., $X$, and stochastic processes are denoted by blackboard bold letters, e.g., $\XX:=(X_i)_{i\in\NN}$.

For $P,Q\in\cP(\cX)$ such that $Q\ll P$, i.e., $Q$ is absolutely continuous w.r.t. $P$, we denote the Radon-Nykodim derivative of $P$ w.r.t. $Q$ by $\frac{dP}{dQ}$.
The KL divergence between $P$ and $Q$, with $P\ll Q$, is
$\DKL(P\|Q):=\EE_P\big[\log\frac{\mathrm{d}P}{\mathrm{d}Q}\big]$. 
The MI between $(X,Y)\sim P_{XY}\in\cP(\cX\times \cY)$ is $\sI(X;Y) := \DKL(P_{XY}\|P_X\otimes P_Y)$, where $P_X$ and $P_Y$ are the marginals of $P_{XY}$.
The entropy of a discrete random variable $X\sim P$ is $H(X) := -\EE\left[\log p(X)\right]$.

\subsection{Directed Information and Channel Capacity}\label{subsec:di}
Originally proposed by Massey \cite{massey1990causality}, DI quantifies the amount of information one sequence of random variables causally conveys about another.
\begin{definition}[Directed information]
Let $(X^n,Y^n)\sim P_{X^n Y^n}\in\cP(\cX^n\times\cY^n)$.
The DI from $X^n$ to $Y^n$ is 
\begin{equation}
    \sI(X^n\to Y^n):= \sum_{i=1}^n \sI(X^i;Y_i|Y^{i-1}).
\end{equation}
\end{definition}
For infinite-time horizon, jointly stationary stochastic processes $\XX$ and $\YY$, the DI rate between them is defined as the asymptotic time-averaged DI:
\[
\sI(\XX\to\YY):= \lim_{n\to\infty}\frac{1}{n}\sI(X^n\to Y^n).
\]
Joint stationarity is indeed sufficient for the existence of this limit\cite{CovThom06}.

Both feedforward and feedback capacities of a sequence of channels $\{P_{Y^n\|X^n}\}_{n\in\NN}$, where $P_{Y^n\|X^n}:=\prod_{i=1}^n P_{Y_i|Y^{i-1}X^i}$, are characterized as\footnote{This formula assumes the so-called information stability property (see \cite{dobrushin1963general}).}
\begin{equation}
C = \lim_{n\to\infty}\sup_{P}\frac{1}{n}\sI(X^n\to Y^n),\label{eq:channel_cap}
\end{equation}
with $P=P_{X^n}$ for feedforward capacity and $P=P_{X^n\|Y^{n-1}}:=\prod_{i=1}^n P_{X_i|X^{i-1}Y^{i-1}}$ (which is termed the causal conditioned distribution) when feedback is present. Also note that when no feedback is present, we have $\sI(X^n;Y^n)=\sI(X^n\to Y^n)$ \cite{massey1990causality}.

\subsection{Directed Information Neural Estimation}
The DINE \cite{tsur2022neural} is an RNN-based estimator of $\sI(\XX\to\YY)$ from a sample $\Dn:=(X^n,Y^n)\sim P_{X^nY^n}$.
Its derivation begins with a representation of DI rate as the asymptotic difference of the following KL divergence terms:
\begin{align*}
    \sD_{Y\|X}^{N}&:=
    \DKL\left(P_{Y^0_{-(N-1)}\| X^0_{-(N-1)}}\middle\| P_{Y^{-1}_{-(N-1)}\| X^{-1}_{-(N-1)}}\otimes P_{\widetilde{Y}}\middle|P_{X^{0}_{-(N-1)}\|Y^{-1}_{-(N-1)}}\right),\\
    \sD_{Y}^{N}&:=\DKL\left(P_{Y^0_{-(N-1)}}\middle\| P_{Y^{-1}_{-(N-1)}}\otimes P_{\widetilde{Y}}\right)\numberthis{}{}\label{eq:DI_rep_kls},
\end{align*}
where $\DKL(P_{Y|X}\|Q_{Y|X}|P_X):=\EE_{P_x}[\DKL(P_{Y|X}\|Q_{Y|X})]$.

The estimator utilizes the Donsker-Varadhan (DV) variational form of the KL divergence  \cite[Theorem 3.2]{donsker1983asymptotic}, whereby for any $P, Q\in\cP(\cX)$ with $P\ll Q$, we have
\begin{equation}
    \DKL\left(P \middle\| Q\right) = \sup_{f: \cX \to \mathbb{R}}\mathbb{E}_P\left[ f \right] -\log\left(\mathbb{E}_Q[ e^{f} ]\right)\label{eq:DV}.
\end{equation}
The supremum is taken over all measurable functions $f$ for which expectations are finite (termed DV potentials) and is achieved by $f^\star:=\log\frac{d P}{d Q}+c$, for any $c\in\RR$.
Applying the DV formula \eqref{eq:DV} to the KL divergences in \eqref{eq:DI_rep_kls}, we have
\begin{align}
    \sD_{Y\|X}^{N} &= \sup_{f_{xy,N}:\cX^N\times \cY^N\mapsto\RR} \EE
    \left[f_{xy,N}(Y^0_{-(N-1)}, X^0_{-(N-1)})\right] - \log\EE
    \left[e^{f_{xy,N}(\tilde{Y},Y^{-1}_{-(N-1)},X^{-1}_{-(N-1)})}\right]\label{eq:DV_xy_expectation},\\
    \sD_{Y}^{N} &= \sup_{f_{y,N}: \cY^N\mapsto\RR} \EE
    \left[f_{y,N}(Y^0_{-(N-1)})\right] - \log\EE
    \left[e^{f_{y,N}(\tilde{Y},Y^{-1}_{-(N-1)})}\right], \label{eq:DV_y_expectation}
\end{align}
where the supremum-achieving DV potentials of \eqref{eq:DV_xy_expectation} and \eqref{eq:DV_y_expectation} are respectively given by
\begin{equation}\label{eq:f_star_dv}
    f^\star_{y,N} :=\log\frac{p_{Y_0|Y^{-1}_{-N}}}{p_{\tilde{Y}}},\qquad f^\star_{xy,N}:=\log\frac{p_{Y_0|X^0_{-N}Y^{-1}_{-N}}}{p_{\tilde{Y}}}.
\end{equation}
The optimal DV potentials \eqref{eq:f_star_dv} are then approximated by RNNs and expectations are estimated by sample means over jointly distributed input-output sequences. We first define the class of RNNs \cite{jin1995universal}.
\begin{definition}[RNN function class]\label{def:RNN_function_class}
Fix $k,d_i,d_o\in\NN$. The class $\GRNN^{(d_i,d_o,k)}$ of RNNs with $k$ neurons and input-output dimensions $(d_i,d_o)$ is the set of discrete-time, nonlinear systems with the following structure:
\[
 \begin{split}
    s_{t+1} &= -\alpha s_{t} + \mathrm{A}\sigma(s_t+\mathrm{B}x_{t}),\\
    y_t &= \mathrm{C} s_t,
 \end{split} 
 \]
where $s_t\in \RR^k$, $x_t\in\RR^{d_i}$, and $y_t\in\RR^{d_o}$ are, respectively, the state, input, and output (column) vectors, $\mathrm{A}\in\RR^{k\times k}$, $\mathrm{B}\in\RR^{k\times d_i}$, and $\mathrm{C}\in\RR^{d_o\times k}$ are the associated weight matrices, $\alpha\in(-1,1)$ is a fixed constant for controlling state decay, and $\sigma(x)=\frac{1}{1+e^{-x}}$ is the sigmoid function, which acts on vectors component-wise. 
\end{definition}

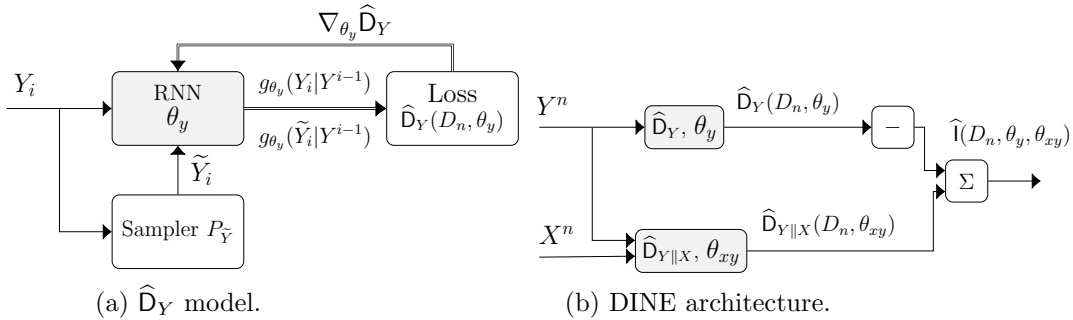
\begin{figure}[t]
    \begin{subfigure}[b]{0.4\textwidth}
        \scalebox{.70}{\usetikzlibrary{arrows.meta, positioning, calc}

\tikzstyle{inv1_block} = [rectangle, minimum width=0.000001cm, minimum height=0.000001cm, draw=none, fill=none]

\tikzstyle{block_D} = [rectangle, rounded corners, minimum width=2.5cm, minimum height=1.4cm, align=center, draw=black, fill= black!5]

\tikzstyle{block_D_w} = [rectangle, rounded corners, minimum width=2.5cm, minimum height=1.4cm, align=center, draw=black, fill= white]

\tikzstyle{block_E} = [rectangle, rounded corners, minimum width=0.8cm, minimum height=0.8cm, align=center, draw=black, fill= black!5]

\tikzstyle{block_C} = [rectangle, minimum width=1cm, minimum height=0.01cm, draw=white, fill=white]
 
\tikzstyle{small_b} = [rectangle, minimum width=0.000001cm, minimum height=0.000001cm, draw=none, fill=black!20]

\begin{tikzpicture}[]

\node (enc) [inv1_block] at (-20,0) {};

\node [block_D, right=2cm of enc] (sampler)  {\large RNN\\ \Large $\theta_y$};

\draw [-{Latex[width=3mm]}] (enc) -- node[above left =1mm and 3mm] {\Large $Y_i$} (sampler);
 
\node [block_D_w, below right = 1.5cm and 2cm of enc] (reference)  {\large Sampler $P_{\tilde{Y}}$};

\node (k) [inner sep=0,minimum size=0,right=1cm of enc] {};

\draw [-{Latex[width=3mm]}] (k) |-  (reference.west);


\draw[-{Latex[width=3mm]}] (reference.north) -- node[right = 0.1cm]{\Large $\tilde{Y}_i$}(sampler.south);

\node (loss) [block_D_w, right=2.7cm of sampler] {\Large Loss\\ \large $\hat{\mathsf{D}}_Y(D_n,\theta_y)$};

\draw [-{Implies},double,-{Latex[width=3mm]}] (sampler) -- node[above=1mm] {\large $g_{\theta_y}(Y_i|Y^{i-1})$} (loss);

\draw [-{Implies},double,-{Latex[width=3mm]}] (sampler) -- node[below=1mm] {\large $g_{\theta_y}(\tilde{Y}_i|Y^{i-1})$} (loss);

\draw [-{Implies},double,-{Latex[width=3mm]}] (loss.north) -- node[above left=0.2cm and 1cm] {\Large $\nabla_{\theta_y}\hat{\mathsf{D}}_Y$} +(0,0.5) -| ([yshift = 0.0cm]sampler.north);

\end{tikzpicture}}
        \caption{$\hat{\sD}_Y$ model.}
        \label{fig:dine_dy}
    \end{subfigure}
    \hspace{0.45cm}
    \begin{subfigure}[b]{0.4\textwidth}
        \scalebox{.70}{\usetikzlibrary{automata, positioning, calc}
\usetikzlibrary{arrows.meta, positioning, calc}

\tikzstyle{block_D} = [rectangle, rounded corners, minimum width=2.5cm, minimum height=1.4cm, align=center, draw=black, fill= black!5]

\tikzstyle{block_D_w} = [rectangle, rounded corners, minimum width=2.5cm, minimum height=1.4cm, align=center, draw=black, fill= white]

\tikzstyle{block_E} = [rectangle, rounded corners, minimum width=0.8cm, minimum height=0.8cm, align=center, draw=black, fill= black!5]

\tikzstyle{block_E_w} = [rectangle, rounded corners, minimum width=0.8cm, minimum height=0.8cm, align=center, draw=black, fill= white]

\tikzstyle{block_C} = [rectangle, minimum width=1cm, minimum height=0.01cm, draw=white, fill=white]
 
\tikzstyle{small_b} = [rectangle, minimum width=0.000001cm, minimum height=0.000001cm, draw=none, fill=black!20]

\tikzstyle{inv1_block} = [rectangle, minimum width=0.000001cm, minimum height=0.000001cm, draw=none, fill=none]

\begin{tikzpicture}[]

\node (Y_in) [inv1_block] at (-1,0) {};

\node [inv1_block, below = 2.2cm of Y_in] (X_in) {};

\node [block_E, right=2cm of Y_in] (D_y)  {\large $\hat{\mathsf{D}}_Y$, \Large $\theta_y$};

\node [block_E, below right=1.8cm and 1.82cm of Y_in] (D_xy)  {\large $\hat{\mathsf{D}}_{Y\|X}$, \Large $\theta_{xy}$};



\node [block_E_w, below right=0.2cm and 4.2cm of D_y] (sum)  {\large $\Sigma$};

\node [block_E_w, right= 2.8cm of D_y] (minus)  {\large $-$};

\node (k) [inner sep=0,minimum size=0,right=1cm of Y_in] {};

\draw [-{Latex[width=3mm]}] (Y_in) -- node[above left =1mm and 3mm] {\Large $Y^n$} (D_y);

\draw [-{Latex[width=3mm]}] (k) |-  ([yshift = 0.2cm]D_xy.west);

\draw [-{Latex[width=3mm]}] (X_in) -- node[above left =1mm and 1mm] {\Large $X^n$} ([yshift = -0.1cm]D_xy.west);

\draw [-{Latex[width=3mm]}] (D_xy) -| node[above right =1mm and -34mm] {\large$\hat{\mathsf{D}}_{Y\|X}(D_n,\theta_{xy})$} +(4.6,1) |- ([yshift = -0.2cm]sum.west);



\draw [-{Latex[width=3mm]}] (D_y) -- node[above right =1mm and -13mm] {\large$\hat{\mathsf{D}}_{Y}(D_n,\theta_{y})$} (minus.west);



\draw [-{Latex[width=3mm]}] (minus) -| +(0.6,-0.8) |- ([yshift = 0.2cm]sum.west);

\node [inv1_block, right = 1cm of sum] (DINE_out) {};

\draw [-{Latex[width=3mm]}] (sum) -- node[above right =5mm and -13mm] {\large$\hat{\mathsf{I}}(D_n,\theta_y,\theta_{xy})$} (DINE_out);

\end{tikzpicture}
        }
        \caption{DINE architecture.}
    \label{fig:dine_complete}
    \end{subfigure}
    
    \caption{The DINE model. Figure (a) depicts a single KL divergence estimator implementation and Figure (b) presents the complete DINE system.}
    \label{fig:dine_figs}
\end{figure}

Note that $\GRNN^{(d_{\mathsf{i}},d_{\mathsf{o}},k)}$ is a parametric class whose (finitely many) parameters belong to some parameter space $\Theta\subset\RR^d$, for an appropriate dimension $d$. When $k$ is fixed, we interchangeably denote functions from the above class explicitly as $g\in\cG_k^{(d_{\mathsf{i}},d_{\mathsf{o}})}$, or in their parametrized form $g_\theta$, where $\theta\in\Theta$. With this notation, the DINE objective is given by \cite{tsur2022neural}
\begin{equation}\label{eq:dine_obj}
\dine(\Dn,\theta_y,\theta_{xy})
    :=\sup_{\theta_{xy}\in\Theta_{xy}}\hat{\sD}_{Y\|X}(\Dn,\theta_{xy})-\sup_{\theta_y\in\Theta_y}\hat{\sD}_{y}(\Dn,\theta_y),
\end{equation}
where $\theta_y\in\Theta_y$ and $\theta_{xy}\in\Theta_{xy}$ are the parameters of the RNNs $g_{\theta_y}\in\GRNN^{(d_y,1,k)}$ and $g_{\theta_{xy}}\in \GRNN^{(d_y+d_x,1,k)}$, respectively, the KL divergence estimators are given by
\begin{subequations}
\begin{align}
   \hat{\sD}_Y(\Dn, \theta_y) &:= \frac{1}{n}\sum_{i=1}^n{g_{\theta_y}}\left(Y^i\right)-\log\left(\frac{1}{n}\sum_{i=1}^n e^{g_{\theta_y}\left(\widetilde{Y}_i,Y^{i-1}\right)}\right),\\
   \hat{\sD}_{Y\|X}(\Dn,g_{\theta_{xy}}) &:= \frac{1}{n}\sum_{i=1}^n{g_{\theta_{xy}}}\left(Y^i,X^i\right)-\log\left(\frac{1}{n}\sum_{i=1}^n e^{g_{\theta_{xy}}\left(\widetilde{Y}_i,Y^{i-1},X^i\right)}\right),
\end{align}\label{eq:DINE_KL_est_main}%
\end{subequations}
and $\tilde{Y}^n$ is an i.i.d. sequence drawn  from $\mathrm{Unif}(\cY)$.
The DINE is now given by optimizing both KL estimators over the corresponding parameter spaces:
\begin{align*}
\dine(\Dn)
    &:=\sup_{\theta_{xy}\in\Theta_{xy}}\inf_{\theta_{y}\in\Theta_{y}}\dine(\Dn,\theta_y,\theta_{xy})\\
    &=\sup_{\theta_{xy}\in\Theta_{xy}}\hat{\sD}_{Y\|X}(\Dn,\theta_{xy})-\sup_{\theta_y\in\Theta_{Y}}\hat{\sD}_{y}(\Dn,\theta_y),
\end{align*}
The DINE architecture is portrayed in Figure \ref{fig:dine_figs}.
For formal consistency guarantees for DINE, as well as implementation details, the reader is referred to \cite{tsur2022neural}.

\subsection{Markov Decision Processes}\label{sec:mdp_def}
MDPs are discrete-time stochastic control processes that are used for sequential decision-making in stochastic systems \cite{bertsekas1976dynamic}. An MDP is described by a tuple $(\cZ,\cU,\cW,P_{W|Z,U},f,r)$, where $\cZ$ and $\cU$ are the state and action spaces, respectively, $W\sim P_{W|Z,U}(\cdot|z,u)\in\cP(\cW)$ is the disturbance given $(z,u)\in\cZ\times\cU$, $r:\cU\times\cZ\to\RR$ is the immediate reward, and the function $f:\cZ\times\cU\times\cW\to\cZ$ describes the state evolution, i.e., $z_{t+1} = f(z_t,u_t,w_t)$.
The action is determined by the stochastic policy $\bm{\pi}=(\pi_t)_{t\in\NN}$, where each $\pi_t$ is a conditional distribution of $U_t$ given $Z_t$, i.e., if $Z_t=z$ then  $U_t\sim\pi_t(\cdot|z)\in\Delta_{\cU}$, for each $t\in\NN$.

We consider an \textit{infinite-horizon average-reward} MDP, where the objective is given by
\begin{equation}\label{eq:avg_reward}
    \rho(\bm{\pi}) := \lim_{N\to\infty}\frac{1}{N}\sum_{t=1}^N \EE_{\bm{\pi}}\left[r(U_t,Z_t)\right],
\end{equation}
where the subscript $\bm{\pi}$ emphasizes that it induces the sequence distribution.
The goal of an MDP agent is to find a policy $\bm{\pi}$ 
that maximizes $\rho(\bm{\pi})$. 
While a priori an optimizing $\bm{\pi}$ may be arbitrary, it turns out that for any infinite-horizon MDP, if $|\cZ|<\infty$, then $\argmax_{\bm{\pi}}\rho(\bm{\pi})$ contains a stationary policy \cite{bertsekas1976dynamic}, i.e., such that $\pi_t(\cdot|z)=\pi(\cdot|z)$ for some conditional distribution $\pi$ and all $z\in\cZ$ and $t\in\NN$.


\section{Directed Information Optimization}\label{sec:di_opt}

We develop a new method for optimizing the DINE over discrete input spaces, leveraging an RNN-based generative model of the input process PMF. To arrive at a tractable optimization objective, we formulate the DI rate optimization problem as an MDP and invoke the policy gradients theorem \cite{sutton1999policy} along with function approximation results and MC methods. 

Henceforth, $\XX$ and $\YY$ denote jointly stationary discrete-time stochastic processes whose samples take values in $\cX$ and $\cY$, respectively. Although applicable in general, our method is presented in the context of communication channels, where $\cX$ and $\cY$ are interpreted as the channel input and output spaces, respectively. We assume that the size of the input alphabet $|\cX|=k<\infty$ is known. For simplicity of presentation, we focus on the case where $\cY$ is also finite, and the channel is described by the causally conditional PMF $p_{Y^n\|X^n}$.
Nonetheless, our derivation is independent of the output alphabet, and readily extends to channels with continuous outputs. 

\subsection{DI Rate Optimization Problem}
We set up the DI rate optimization problem  (see Section \ref{subsec:di}), modeling the input process PMF by a deep generative model as described next. The generative model takes an input-output pair from $\cX\times\cY$ and a simplex vector (that models the current input PMF), and outputs a new simplex vector (the updated input PMF). The PMF generator corresponding to a parameter vector $\phi\in\Phi\subset\RR^d$ is denoted by $h_\phi:\cX\times\cY\times\Delta_k\to\Delta_k$. Since each new simplex vector $p_t^\phi$ is specified by the parameters and the sequence of past input-output pairs $(X^{t-1},Y^{t-1})$, we treat the $t$-th output of $\pmf$ as the model for the conditional PMF of $X_t$ given that past: 
\[
p_t^\phi=p^\phi(\cdot|X^{t-1},Y^{t-1}):= h_\phi(X_{t-1}, Y_{t-1},p^\phi_{t-1}),\qquad t\geq 1,
\]
where $(X_s,Y_s)\sim p_s^\phi p_{Y_s|X^sY^{s-1}}$ for each $s\leq t-1$.


The goal is now to optimize the DI rate over all input distributions that are modeled by $\pmf$, $\phi\in\Phi$, i.e., to solve
\begin{equation}\label{eq:di_phi_opt}
    \sup_{\phi\in\Phi}\sI_{\phi}(\XX\to\YY),
\end{equation}
where the subscript $\phi$ designates that the underlying distribution for each $\sI(X^n\to Y^n)$, $n\in\NN$, in the DI rate expression is $\prod_{t=1}^n \pt p_{Y_t|X^{t}Y^{t-1}}$. Note that $\sI_{\phi}(\XX\to\YY)$ exists due to the stationarity of the joint distribution, and when $\Phi$ is compact, the supremum in \eqref{eq:di_phi_opt} is attained. To solve \eqref{eq:di_phi_opt}, we seek a tractable expression for the DI rate gradient $\nabla_\phi\sI_{\phi}(\XX\to\YY)$. The next section reformulates DI rate optimization as an MDP and employs the policy gradients theorem, alongside a DINE-based approximation of the MDP reward to arrive at an objective whose gradients coincide with those of the above.
The section concludes with a deep reinforcement learning policy optimization methodology for the calculation of \eqref{eq:di_phi_opt}.
\begin{table}[t]
\centering
\caption{DI rate optimization MDP formulation}
 \begin{tabular}{||c | c ||} 
 \hline
 MDP & DI optimization  \\ 
 \hline \hline
 State \vspace{0.02cm}$Z_t$ & $X^{-1}_{-t},Y^{-1}_{-t}$  \\
 \hline
 Action $U_t$ & $X_0$  \\
 \hline
 Disturbance $W_t$ & $Y_0$  \\
 \hline
 Reward $r(U_t,Z_t)$ & Eqn. \eqref{eq:mdp_reward}  \\
 \hline \hline
\end{tabular}
\label{table:DP}
\end{table}

\subsection{MDP Formulation}

Recall that an MDP is given by the tuple $(\cZ,\cU,\cW,P_{W|Z,U},f,r)$. By the stationarity of the model, we may apply a reverse time-shift operator on each time step, so that the most recent step remains $t=0$ throughout. To obtain an MDP formulation of the DI rate optimization, we take the state as the accumulation of past channel inputs and outputs, i.e., $Z_t=(X^{-1}_{-t},Y^{-1}_{-t})$.\footnote{To ensure that the state space is the same for all $t$, we concatenate the state $Z_t=(X^{-1}_{-t},Y^{-1}_{-t})$ with infinitely many null symbols such that $\cZ$ is a space of half-infinite sequences.}
We view the channel input generator as an agent whose action $U_t=X_0$ at each step is drawn from the parametric policy $\pi_\phi(\cdot|Z_t)=p^\phi_{t}(\cdot|X^{-1}_{-t},Y^{-1}_{-t})$.
The disturbance is the channel output, distributed according to the conditional PMF $p_{Y_0|Y^{-1}_{-t}X^0_{-t}}$, and the immediate reward is given by the conditional expectation
\begin{equation}
    r(U_t,Z_t) = \EE\left[\log\left(\frac{p_{Y_0|Y^{-1}_{-t},X^{0}_{-t}}(Y_0|Y^{-1}_{-t},X^{0}_{-t})}{p_{Y_0|Y^{-1}_{-t}}(Y_0|Y^{-1}_{-t})}\right)
    \middle|X^{0}_{-t},Y^{-1}_{-t}\right],\label{eq:mdp_reward}
\end{equation}
for which we have $\EE\left[r(U_t,Z_t)\right] = \sI_\phi(X^0_{-t};Y_0|Y^{-1}_{-t})$. For feedforward communication, the MDP formulation remains unchanged, while the optimization is limited to policies that are independent of past channel outputs. The formulation is summarized in Table \ref{table:DP}, and gives rise to the desired MDP characterization (see Section \ref{proof:mdp_formulation} for the proof).

\begin{theorem}[MDP formulation]
\label{theorem:MDP_formulation}
The DI rate optimization problem \eqref{eq:di_phi_opt} is an infinite-horizon average-reward MDP with objective $\rho(\pi_\phi) = \sI_\phi(\XX\to \YY)$.
\end{theorem}

\begin{remark}[Relation to existing MDPs]
MDP formulations of capacity-optimization problems were previously used to calculate the feedback capacity of certain unifilar FSCs \cite{permuter2008capacity,elishco2014capacity,sabag2015feedback}, assuming the channel model is known.
In these formulations the MDP state space is a quantized version of $\Delta_{|\cS|}$, and $\rho(\pi)$ is a single-letter expression, which is optimized using dynamic programming algorithms. 
The authors of \cite{tatikonda2008capacity} generalize the unifilar formulation but obtain a generally intractable objective.  
Our formulation, on the other hand, can be viewed as a unifying MDP for all channels that have a stationary joint distribution. As in \cite{tatikonda2008capacity}, our MDP cannot be computed with traditional methods and calls for new ideas, utilizing approximation and estimation techniques. 
\end{remark}


\begin{remark}[Existence of optimal solution]
    Optimal policies are guaranteed for MDPs with finite state and action spaces \cite{bertsekas1976dynamic}. However, when the state space becomes infinite, an optimal policy is no longer guaranteed unless the MDP satisfies certain conditions on its ergodicity (cf. \cite{cavazos1992comparing,arapostathis1993discrete,ross2014introduction,xia2020existence}).
    This is the case for prior methodologies that used MDPs to maximize DI, where the MDP state space is a probability simplex \cite{permuter2008capacity,elishco2014capacity,sabag2015feedback, aharoni2020reinforcement, shemuel2021feedback}.
\end{remark}

%

%

\subsection{Policy Gradients Theorem}
We leverage the MDP formulation to arrive at a tractable expression for $\nabla_\phi \rho(\pi_\phi)$ using reinforcement learning techniques. This approach is particularly well-suited here since we treat the channel as a black-box that can be sampled given an input sequence, and reinforcement learning offers powerful tools for solving data-driven MDPs.
We invoke the policy gradients theorem \cite[Theorem 1]{sutton1999policy}, that enables expressing $\nabla_\phi \rho(\pi_\phi)$ in terms of $\nabla_\phi \pi_\phi$, which is typically simpler to compute.

\begin{theorem}[Policy gradients]\label{theorem:policy_grads}
Let $(\cZ,\cU,\cW,P_{W|Z,U},f,r)$ be an infinite-horizon average-reward MDP with objective $\rho$, and consider a parameterized stationary policy $\pi_\phi$, where $\phi\in\Phi\subseteq\RR^d$ for some $d\in\NN$. Define the function
\begin{equation}
    \sQ^{\pi_\phi}(u,z) :=\sum_{t=1}^\infty\EE\left[r(U_t,Z_t) - \rho(\pi_\phi)\big{|}Z_0=z,U_0=u\right].\label{eq:q_func_def}
\end{equation}
Then,
\begin{equation}\label{eq:policy_grad_thm}
    \nabla_\phi\rho(\pi_\phi) = \sum_{z\in\cZ}p_Z^{\pi_\phi}(z)\sum_{u\in\cU}\nabla_\phi\pi_\phi(u,z) \sQ^{\pi_\phi}(u,z),
\end{equation}
where $p_Z^{\pi_\phi}$ is the stationary distribution of the MDP state sequence.
\end{theorem}

The state-action value function $\mathsf{Q}^{\pi_\phi}$ in \eqref{eq:q_func_def} quantifies the expected deviation of future rewards from $\rho(\pi_\phi)$, for a given state-action pair. The policy gradients theorem simplifies $\nabla_\phi \rho(\pi_\phi)$ by representing it in terms of the policy gradient $\nabla_\phi\pi_\phi$ and the function $\mathsf{Q}^{\pi_\phi}$.  Using the identity $\partial_x \log f(x) =  \frac{\partial_x f(x)}{f(x)}$ in \eqref{eq:policy_grad_thm}, we may further represent
\begin{align*}
    \nabla_\phi\rho(\pi_\phi)
    &= \sum_{z}p_Z^{\pi_\phi}(z)\sum_{u}\pi_\phi(u,z)\nabla_\phi\log\big(\pi_\phi(u,z)\big) \sQ^{\pi_\phi}(u,z)\\
    &= \EE\left[ \nabla_\phi\log\big(\pi_\phi(U,Z)\big) \sQ^{\pi_\phi}(U,Z)  \right].\numberthis{}\label{eq:log_der_identity}
\end{align*}
We next use the DINE to approximate $\sQ^{\pi_\phi}$ and arrive at a tractable expression for \eqref{eq:log_der_identity}.

\subsection{Approximation via DINE}
To compute the desired gradient via the right-hand side (RHS) of \eqref{eq:log_der_identity}, one must evaluate the function $\sQ^{\pi_\phi}$. This, however, requires calculation of the MDP reward \eqref{eq:mdp_reward}, which necessitates knowledge of the channel model. Even if the channel is given, traditional tabular algorithms cannot be efficiently applied due to the size of the MDP state space. We circumvent this by using the DINE \cite{tsur2022neural} to approximate $\sQ^{\pi_\phi}$ based only on samples from the channel.

Fix the PMF generator parameters $\phi\in\Phi$ and consider a dataset drawn from this input PMF and the channel $\Dn = (X^{n},Y^n)\sim \prod_{t=1}^n p^\phi_t p_{Y_t|Y^{t-1}X^t}$. We first take the DINE $\dine(\Dn)$ as a strongly consistent estimate of the true DI rate $\rho(\pi_\phi)$, cf. \cite[Theorem~2]{tsur2022neural}. Then, the function $r(U_t,Z_t)$ in the definition of $\sQ^{\pi_\phi}$ is approximated using the trained DINE RNNs $(g_{\theta_{y}}, g_{\theta_{xy}})$, as follows.
We consider the DV representation of the derived KL divergences in \eqref{eq:DV_xy_expectation} and \eqref{eq:DV_y_expectation}.
Subtracting their supremum-achieving DV potentials \eqref{eq:f_star_dv} yields the likelihood ratio 
$f^\star_{xy,N} - f^\star_{y,N} = \log\left(\frac{p_{Y_0|X^0_{-N}Y^{-1}_{-N}}}{p_{Y_0|Y^{-1}_{-N}}}\right)$, where
\[
    \EE[r(U_N,Z_N)] = \EE\left[f^\star_{xy,N}(X^0_{-N},Y^{0}_{-N}) - f^\star_{y,N}(Y^0_{-N})\right].
\]
As the DINE RNNs $(g_{\theta_{xy}},g_{\theta_{y}})$ are optimized to achieve the supremum of the empirical DV forms corresponding to \eqref{eq:DV_xy_expectation} and \eqref{eq:DV_y_expectation}, we define $\hat{r}_{\theta} := g_{\theta_{xy}} - g_{\theta_y}$, and take it as a proxy for $r$. This construction assumes that $\phi\in\Phi$ is fixed and that $g_{\theta_y}$ and $g_{\theta_{xy}}$ have been optimized for the induced joint distribution $\prod_{t=1}^n p^\phi_t p_{Y_t|Y^{t-1}X^t}$. This assumption will be further discussed in Section \ref{sec:implementation}, where the joint optimization algorithm is proposed.

For the last step in approximating $\sQ^{\pi_\phi}$, we observe that 
\[
    \lim_{t\to\infty}\EE\left[r(U_t,Z_t)\right] = \lim_{t\to\infty}\sI_{\phi}(X^0_{-t};Y_0|Y^{-1}_{-t})= \sI_\phi(\XX\to\YY)=\rho(\pi),
\]
which implies that the difference $r(U_t,Z_t)-\rho(\pi_\phi)$ becomes negligible as $t$ grows. This serves to justify truncating the infinite sum defining $\sQ^{\pi_\phi}$ at some $T<\infty$, which, together with the steps above, yields the approximation 
\[
     \hat{\sQ}_{\theta,t}(\Dn) := \sum_{i=t}^{t+T-1}
     \hat{r}_\theta(Y^{i},X^{i})-\hat{\sI}(\Dn,\theta).
\]

Having this, we estimate the RHS of \eqref{eq:log_der_identity} by replacing the outer expectation with a MC evaluation taken over a long trajectory $(\pt, X_t, Y_t)_{t=1}^n$, which results in 
\begin{equation}
     \nabla_\phi\underbrace{\frac{1}{n-T}\sum_{t=1}^{n-T}\log \left(p^\phi_t(X_t)\right)\hat{\sQ}_{\theta,t}(\Dn)}_{=:\hat{\sJ}_\theta(\Dn,\phi)}\label{eq:alt_objective}
\end{equation}
as the final approximation of $\nabla_\phi\rho(\pi_\phi)$. This objective is readily differentiable w.r.t. $\phi$ based only on samples from the input generative model and the channel, as desired.
Consequently, the DI optimization scheme alternates between optimizing $\hat{\sI}(\Dn)$ and $\hat{\sJ}_\theta(\Dn,\phi)$, i.e., alternates between improving the approximation of $\hat{r}_\theta$ and policy optimization, respectively.


\begin{remark}[Performance guarantees]
    While the DINE is a consistent estimator of DI \cite{tsur2022neural}, formal guarantees for our overall method would require a theoretical account of RNN-based policy optimization combined with MC schemes, which is currently unavailable \cite{lillicrap2015continuous}. Nevertheless, in Section \ref{sec:exp_cap} we show empirically that our approach performs well on all the considered examples. 
\end{remark}

\subsection{Estimated Mutual Information Optimizer}\label{sec:mine_opt}
When the channel is memoryless the optimization problem \eqref{eq:di_phi_opt} reduces to maximizing $\sI_\phi(X;Y)$ over input generative models that are elements of the simplex $\Delta_k$.
We can take advantage of the memoryless structure in this special case and employ a simpler model. 
First, $\pmf$ no longer needs to depend on past channel input-outputs pairs and can be taken as $p^\phi=\big(\softmax^1(\phi),\ldots,\softmax^m(\phi)\big)$,~where 
\begin{equation}\label{eq:softmax_def}
    \softmax^i(\phi) := \frac{\exp(\phi_i)}{\sum_{j=1}^m\exp(\phi_j)}, \qquad i=1,\dots,m
\end{equation}
is the $i$-th output of the softmax function. 

Second, we replace the DINE with the MI neural estimator (MINE) \cite{belghazi2018mutual}:
\begin{equation}
    \hat{\sI}_{\mathsf{MI}}(\Dn) := \sup_{\theta\in\Theta}\frac{1}{n}\sum_{t=1}^n g_\theta(X_i,Y_i) - \log\left( \frac{1}{n}\sum_{t=1}^n e^{g_\theta(X_t,\bar{Y}_t)}\right),
\end{equation}
where $g_\theta$ is a feedforward neural network, $\Dn=(X^n,Y^n)\stackrel{i.i.d.}{\sim}p^\phi p_{Y|X}$, and $\bar{Y}_t$ are negative samples that are independent of $X_t$ for $t=1,\dots n$. 
MINE is optimized over feedforward networks, which are simpler and often present better convergence profiles. 
In addition, MINE lower bounds the ground truth MI \cite[Remark~5]{tsur2022neural} (which is not guaranteed for DINE) and adheres to non-asymptotic error bounds \cite{sreekumar2021non,sreekumar2022neural}.
In Section \ref{sec:prob_shape} we demonstrate the MI-based optimization scheme to learn probabilistic shaping of constellations in the peak-power constrained AWGN (PP-AWGN).

The resulting differentiation objective takes the form:
\begin{equation}
    \hat{\sJ}^{\mathsf{MI}}_\theta(\Dn,\phi) := \frac{1}{n}\sum_{t=1}^{n}\log\left( p^\phi(X_t)\right)\left(g_\theta(X_t, Y_t) - \hat{\sI}_{\mathsf{MI}}(\Dn)\right),\label{eq:alt_objective_mine}
\end{equation}
whose gradient can be simplified as follows (see Section \ref{proof:mine_based_grad} for the proof).
\begin{lemma}\label{lemma:MINE_grad}
The gradient of \eqref{eq:alt_objective_mine} w.r.t. $\phi$ is given by
\begin{equation}
\nabla_\phi\hat{\sJ}^{\mathsf{MI}}_\theta(\Dn,\phi) = \frac{1}{n}\sum_{t=1}^{n}\left(e_{X_t} -  p^\phi\right)\left(g_\theta(X_t, Y_t) - \hat{\sI}_{\mathsf{MI}}(\Dn)\right),
\end{equation}
where $e_{X_t}=\big(\mathbbm{1}_{\{X_t=1\}}\ldots\mathbbm{1}_{\{X_t=m\}}\big)^\intercal$ is the $m$-dimensional standard basis vector whose $X_t$-th entry is 1.
\end{lemma}

\section{Implementation and Algorithm}\label{sec:implementation}
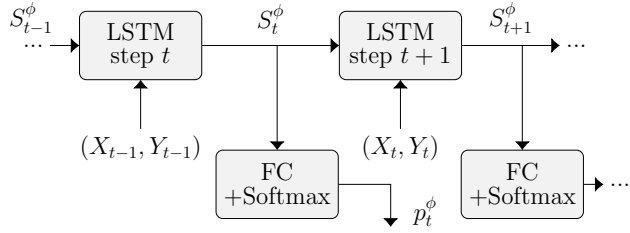
\begin{figure}[t]
    \centering
     \scalebox{.65}{\usetikzlibrary{arrows.meta, positioning, calc}

\tikzstyle{block_D} = [rectangle, rounded corners, minimum width=2.5cm, minimum height=1.4cm, align=center, draw=black, fill= black!5]

\tikzstyle{block_E} = [rectangle, rounded corners, minimum width=0.8cm, minimum height=0.8cm, align=center, draw=black, fill= black!5]

\tikzstyle{block_C} = [rectangle, minimum width=1cm, minimum height=0.01cm, draw=white, fill=white]
 
\tikzstyle{small_b} = [rectangle, minimum width=0.000001cm, minimum height=0.000001cm, draw=none, fill=black!20]

\begin{tikzpicture}

\node (LSTM_t_i) [block_D] at (-1,0) {\Large LSTM\\\Large step $t$};
\node [block_D, right=2.8cm of LSTM_t_i] (LSTM_t_i_1)  {\Large \Large LSTM\\\Large step $t+1$};


\node[below=2cm of LSTM_t_i] (dummy_i) {}; 
\node[below=2cm of LSTM_t_i_1] (dummy_i_1) {}; 

\node (Preprocess_i) [block_D, right= 1.4cm of dummy_i] {\Large FC \\ \Large +Softmax};
\node (Preprocess_i_1) [block_D, right= 1.1cm of dummy_i_1] {\Large FC \\ \Large +Softmax};
\draw [-{Latex[width=3mm]}] (LSTM_t_i) -- node[above=1mm] {\Large $S_t^\phi$} (LSTM_t_i_1);


\node [right=2cm of LSTM_t_i_1] (dummy_LSTM_i_1)  {\Large ...};
\draw [-{Latex[width=3mm]}] (LSTM_t_i_1) -- node[above=1mm] {\Large $S_{t+1}^\phi$} (dummy_LSTM_i_1);

\draw [-{Latex[width=3mm]}] (LSTM_t_i.east) -- +(0,0) -| (Preprocess_i.north);
\draw [-{Latex[width=3mm]}] (LSTM_t_i_1.east) -- +(0,0) -| (Preprocess_i_1.north);

\node (dummy_h_i) [below =1cm of LSTM_t_i]  {\Large $(X_{t-1},Y_{t-1})$}; 
\node (dummy_h_i_1) [below =1cm of LSTM_t_i_1]  {\hspace{-0.5cm}\Large $\quad(X_t,Y_t)$}; 

\draw [-{Latex[width=3mm]}] (dummy_h_i.north) -- +(0,0) -| (LSTM_t_i.south);
\draw [-{Latex[width=3mm]}] (dummy_h_i_1.north) -- +(0,0) -| (LSTM_t_i_1.south);

\node (dummy_p_out_i) [below =3cm of LSTM_t_i_1]  {};
\draw [-{Latex[width=3mm]}] (Preprocess_i.east) --node[below right = 0.1cm and 1.4cm] {\Large $p^\phi_t$} +(0,0) -| ([xshift=-0.2cm]dummy_p_out_i.north);

\node (dummy_p_out_i_1) [right = 0.4cm of Preprocess_i_1]  {\Large ...};
\draw [-{Latex[width=3mm]}] (Preprocess_i_1.east) --  (dummy_p_out_i_1);

\node (dummy_past_s) [left = 0.6cm of LSTM_t_i]  {\Large ...};
\draw [-{Latex[width=3mm]}] (dummy_past_s) --node[above left = 1mm and 1mm] {\Large $S_{t-1}^\phi$} (LSTM_t_i);
\end{tikzpicture}}
    \caption{The PMF model unrolled for feedback capacity. In the $t$-th step, $S^\phi_t$ is calculated from $(S^\phi_{t-1},X_t, Y_t)$ and then passed for the calculation of both $S^\phi_{t+1}$ and $\pt$.
    }
    \label{fig:pmf_gen}
\end{figure}

\subsection{PMF Generator}
Recall that the PMF generator calculates a mapping $h_\phi:\cX\times\cY\times\Delta_k\to\Delta_k$, whose output evolves according to $p^\phi_t = h_\phi(X_{t-1},Y_{t-1},p^{\phi}_{t-1})$.
We implement $h_\phi$ with a long short-term memory (LSTM) network.
LSTM is a type of recurrent neural network that uses gating mechanisms to selectively retain, input, output, and forget information over multiple time steps, allowing it to model long-term dependencies in sequential data (see \cite{hochreiter1997long} for more background on LSTMs).
We stack the LSTM network with additional fully-connected (FC) networks to increase the expressiveness of the model. The output layer of $h_\phi$ is given by an $m$-dimensional softmax activation \eqref{eq:softmax_def}.
Denoting the LSTM and FC maps by $g^\phi_1$ and $g^\phi_2$, respectively, the PMF generator output evolution is given by
\begin{equation}
    S^\phi_t = g^\phi_1(X_{t-1},Y_{t-1},S^{\phi}_{t-1})\,,\quad
    p^\phi_t= \sigma_{\text{sm}}\big(g^\phi_2(S^\phi_t)\big),\label{eq:pmf_state_def}
\end{equation}
where $S^\phi_t$ is the LSTM inner state at time $t$.
When feedforward capacity is considered, $Y_{t-1}$ is omitted from \eqref{eq:pmf_state_def}. The architecture of $h_\phi$ is illustrated in Figure \ref{fig:pmf_gen}.

\subsection{Combined System}
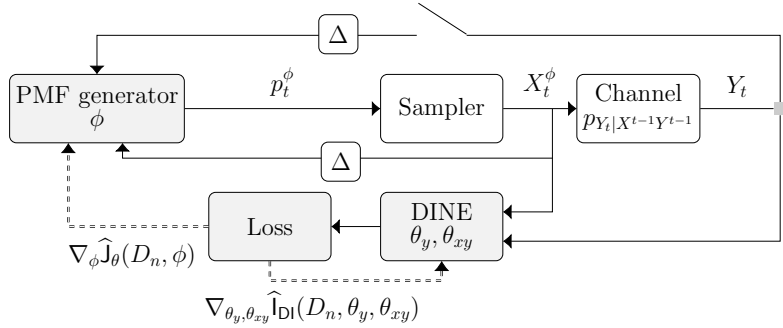
\begin{figure}[t]
    \centering
     \scalebox{.65}{\usetikzlibrary{arrows.meta, positioning, calc}

\tikzstyle{block_D} = [rectangle, rounded corners, minimum width=2.5cm, minimum height=1.4cm, align=center, draw=black, fill= black!5]

\tikzstyle{block_white} = [rectangle, rounded corners, minimum width=2.5cm, minimum height=1.4cm, align=center, draw=black, fill= white!5]

\tikzstyle{block_E} = [rectangle, rounded corners, minimum width=0.8cm, minimum height=0.8cm, align=center, draw=black, fill= white!5]

\tikzstyle{block_C} = [rectangle, minimum width=1cm, minimum height=0.01cm, draw=white, fill=white]
 
\tikzstyle{small_b} = [rectangle, minimum width=0.000001cm, minimum height=0.000001cm, draw=none, fill=black!20]

\tikzstyle{block_l1} = [rectangle, minimum width=0.0cm, minimum height=0.00cm, draw=white, fill=white]

\tikzstyle{block_l0} = [rectangle, minimum width=0.71cm, minimum height=0.01cm, draw=white, fill=white]

\begin{tikzpicture}[]
\node (enc) [block_D] at (-10,0) {\Large PMF generator\\\Large $\phi$};
\node [block_white, right=4cm of enc] (sampler)  {\Large Sampler};
\node (channel) [block_white, right=1.5cm of sampler] {\Large Channel\\ \Large \textbf{$p_{Y_t|X^{t-1}Y^{t-1}}$}};
\node (DINE) [block_D, below=1cm of sampler] {\Large DINE \\ \Large $\theta_y,\theta_{xy}$};
\node (loss) [block_D, left=1cm of DINE] {\Large Loss};
\draw [-{Latex[width=3mm]}] (DINE) -- (loss);

\node (k) [inner sep=0,minimum size=0,right=1cm of sampler] {}; 
\node (k2) [inner sep=0,minimum size=0, below right =0.3cm and 1cm of sampler] {}; 
\node (k3) [inner sep=0,minimum size=0, below right =0.6cm and 1.6cm of channel] {}; 

\draw [-{Latex[width=3mm]}] (enc) -- node[above=1mm] {\Large $p^\phi_t$} (sampler);
\draw [-{Latex[width=3mm]}] (sampler) -- node[above=1mm] {\Large $X^\phi_t$} (channel);

\node [small_b, right=1.5cm of channel] (invisible) {};
\draw (channel) -- node[above=1mm] {\Large $Y_t$} (invisible);

\draw [-{Latex[width=3mm]}] (k2) -- +(0,0) -| ([xshift = 0.75cm]enc);
\draw [-{Implies},double,dashed,-{Latex[width=3mm]}] (loss) --node[below left =0.9mm and 0.1cm] {\hspace{-2cm}\Large $\nabla_\phi \hat{\sJ}_\theta(\Dn, \phi)$} +(-1.3,0) -| ([xshift = -2cm]enc);

\draw [-{Latex[width=3mm]}] (k) -- +(0,-1.7) |- ([yshift = 0.3cm]DINE.east);
\draw[-{Latex[width=3mm]}] (invisible) -- +(0,-1.9)  |- ([yshift = -0.3cm]DINE.east);

\draw[-{Latex[width=3mm]}] (invisible) -- +(0,1.5) -| (enc);
\node (FB_space) [block_C, above =0.68cm of sampler] {};
\node (FB_end) [block_C, above right =0.5cm and -1.7cm of FB_space] {};
\node (FB_delta) [block_E, left = 1.2cm of FB_space] {\Large$\Delta$};
\draw (FB_space.east) -- (FB_end);


\node (X_delta) [block_E, above left = -2.2cm and 0.4cm of sampler] {\Large$\Delta$};

\draw [-{Implies},double,dashed,-{Latex[width=3mm]}] (loss.south) -- node[below left =2.7mm and -3.2cm] {\hspace{-2cm}\Large $\nabla_{\theta_y,\theta_{xy}} \hat{\sI}_{\mathsf{DI}}(\Dn,\theta_y,\theta_{xy})$} +(0,-0.4) -| (DINE.south);


\end{tikzpicture}}
    \caption{The complete estimation-optimization model. Dashed arrows represent gradient propagation and filled blocks represent parametric models.}
    \label{fig:complete_di_system}
\end{figure}
The combined system comprises both the PMF generator and the DINE models, as presented in Figure \ref{fig:complete_di_system}.
We therefore construct a joint estimation-optimization procedure based on alternating maximization between $\dine(\Dn,\theta_y,\theta_{xy})$ and $\hat{\sJ}_\theta(\Dn,\phi)$.
In each iteration, a single model is selected for parameter update, while the parameters of the other are fixed.
Optimizing the DINE improves the approximation accuracy of $\sQ^{\pi_\phi}$ for a fixed input PMF model $h_\phi$, while optimizing $h_\phi$ increases DI, as quantified by the current DINE model.


In each iteration, we perform the following steps:
First, we calculate $p^{\phi,n}$ and a dataset $D_n$ by sequentially calculating the PMFs, sampling from them, and propagating the sampled inputs through the channel.
Then, we pass $D_n$ through the DINE and calculate the loss function.
We update the models' parameters using stochastic gradient ascent;
for the $\theta$ update, we calculate $\nabla_{\theta_y,\theta_{xy}}\dine(\Dn,\theta_y,\theta_{xy})$, while for the $\phi$ update, we calculate $\nabla_\phi\hat{\sJ}_\theta(\Dn,\phi)$.
These steps are repeated until a convergence criterion is met or a predetermined number of updates is performed.
Finally, we evaluate the DINE objective on a long sequence of channel inputs and outputs to obtain a numerical estimate of the optimized DI.
See Algorithm~\ref{alg:di_opt} for the full list of steps. 

The proposed joint optimization method involves a latent assumption; updating the PMF generator requires an accurate estimate of $\sQ^{\pi_\phi}$ w.r.t. the joint distribution
induced by the current value of $\phi$.
We therefore prioritize the training of the DINE model and apply several updates to the DINE parameters $(\theta_y,\theta_{xy})$ for each update of the PMF generator parameters~$\phi$.

\begin{algorithm}[ht]
\caption{Discrete alphabet DI optimization and estimation}
\label{alg:di_opt}
\textbf{input:} Discrete channel, feedback indicator\\
\textbf{output:} $\dine(\Dn, \NE)$, optimized $h_\phi$.
\algrule
\begin{algorithmic}
\State Initialize $g_{\theta_{y}}, g_{\theta_{xy}}, \NE$ with parameters $\theta_{y}$, $\theta_{xy}$ and $\phi$ and set a learning rate $\gamma$.
    \If{feedback indicator}
        \State Add feedback link to $h_\phi$
    \EndIf
\Repeat
\State Compute $(\Dn, p^{\phi,n})$
\If{training DINE}
    \State Compute $\hat{\sD}_{Y}(\Dn, \theta_y)$, $\hat{\sD}_{Y \| X}(\Dn, \theta_{xy})$ according to \eqref{eq:DINE_KL_est_main}
    \State Update DINE parameters: \\
    \hspace{\algorithmicindent}\hspace{\algorithmicindent}$\theta_{y} \leftarrow \theta_{y} + \gamma\nabla_{\theta_{y}}\hat{\sD}_{Y}(\Dn, \theta_{y})$\\
    \hspace{\algorithmicindent}\hspace{\algorithmicindent}$\theta_{xy} \leftarrow \theta_{xy} + \gamma\nabla_{\theta_{xy}}\hat{\sD}_{Y \| X}(\Dn, \theta_{xy})$ 
\Else \hspace{0.15cm}(Train PMF generator)
\State Compute $\hat{\sJ}_\theta(\Dn, \phi)$  according to \eqref{eq:alt_objective}

\State Update PMF generator parameters: \\
\hspace{\algorithmicindent}\hspace{\algorithmicindent}$\phi \leftarrow \phi + \gamma\nabla_{\phi}\hat{\sJ}_\theta(\Dn, \phi)$
\EndIf
\Until{convergence}
\State MC evaluation of  $\dine(\Dn)$ \\
\Return $\dine(\Dn)$ and $h_\phi$
\end{algorithmic}
\end{algorithm}

\section{Application 1: Capacity Estimation}\label{sec:exp_cap}
We employ Algorithm~\ref{alg:di_opt} to estimate the capacity of various channels with memory.
Performance is measured by comparing the optimized DI estimate with known capacity solutions and/or bounds.
Further, we compare the learned PMF structure with known capacity-achieving coding schemes.
Unlike the approach developed herein, all the considered reference methods require full knowledge of the channel law.
Implementation can be found on \href{https://github.com/DorTsur/discrete_di_optimization}{GitHub}.

\begin{remark}[Capacity optimization problem]
The proposed method aims to calculate the supremum of the DI rate w.r.t. a set of input distributions.
However, the capacity expression \eqref{eq:channel_cap} considers the opposite order of limit and supremum, i.e., taking the limit of the sequence of optimized normalized DI.
This order is known to be interchangeable for FSCs \cite{permuter2009finite}, which encompass most models of channels with memory.
\end{remark}


\subsubsection{\texorpdfstring{\underline{Gilbert-Eliott Channel}}{Gilbert-Eliott Channel}}
\begin{figure}[t]
  \begin{subfigure}[ht]{.43\textwidth}
    \centering
     \scalebox{.75}{\begin{tikzpicture}
\node[state] (1) {$Bad$};
\node[state, right=2cm of 1] (2) {$Good$};
\draw{
(1) edge[->,>=stealth,thick,below,bend right=20] node{$g$} (2)
(2) edge[->,>=stealth,thick,above,bend right=20] node{$b$} (1)
(1) edge[->,>=stealth,thick,loop left,left=1] node{$1-g$} (1)
(2) edge[->,>=stealth,thick,loop right,right=1] node{$1-b$} (2)
};
\end{tikzpicture}}
    \caption{}
    \label{fig:ge_markov}
  \end{subfigure}
  \begin{subfigure}[ht]{.57\textwidth}
    \centerline{\includegraphics[trim={10pt 1pt 20pt 20.4pt}, clip, width=\linewidth]{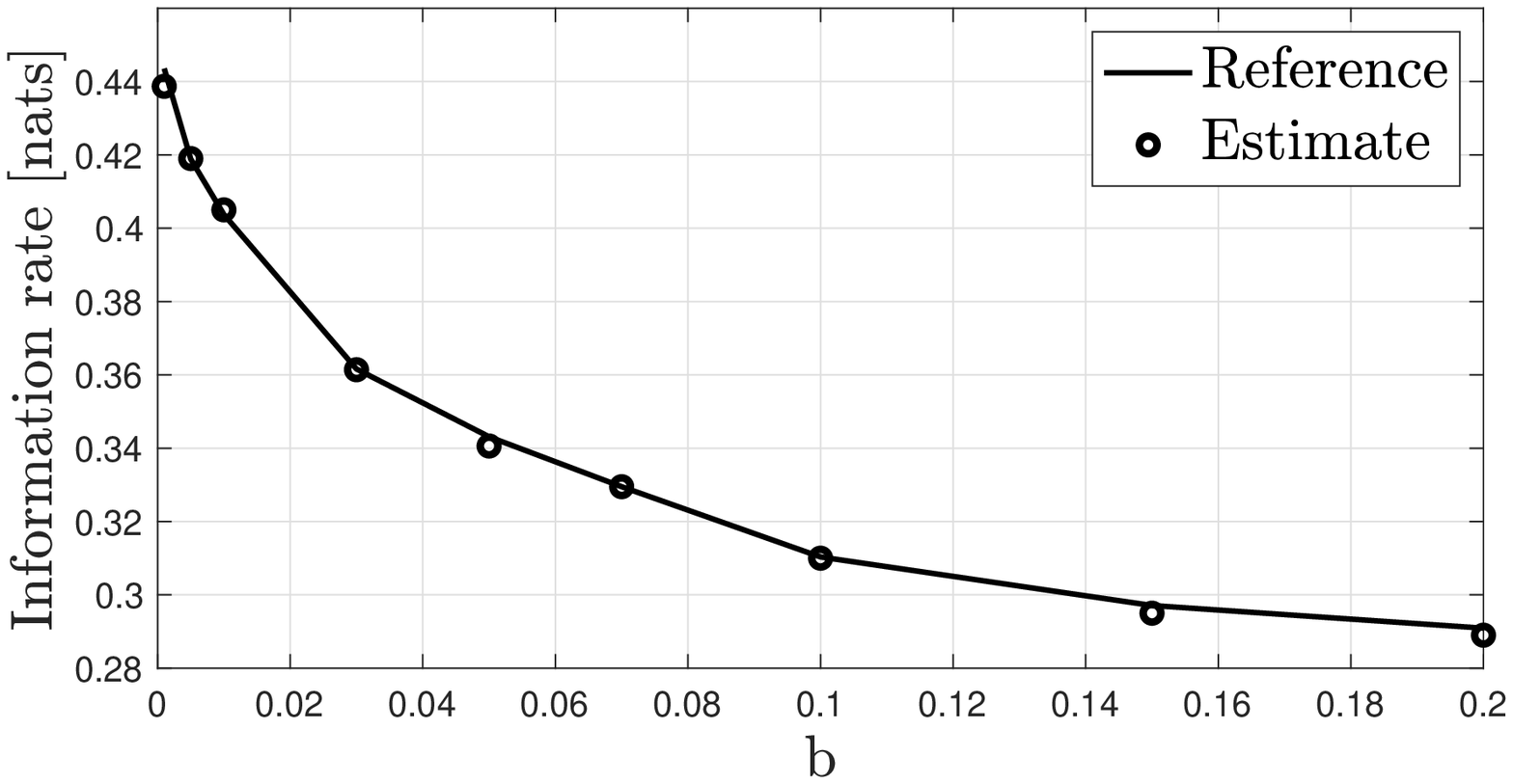}}
    \caption{} 
    \label{fig:GE}
  \end{subfigure}
  
  \caption{GE channel. Figure (a) depicts the channel state Markov chain. The transition distribution from "Good" to "Bad" and vice versa are $\mathrm{Ber}(b)$ and $\mathrm{Ber}(g)$, respectively; (b) presents the estimated capacity versus $b$ (with $g=3b$), compared to estimates obtained from \cite{rezaeian2005computation}.}
\end{figure}

The Gilbert-Eliott (GE) channel is a time-varying binary symmetric channel (BSC) whose flip parameter is determined by a latent Markovian state that evolves according to the state diagram in Figure \ref{fig:ge_markov}. At state `Good' the flip probability is smaller, and thus the channel is better. 
While it is straightforward to show that the optimal input is an i.i.d. $\mathrm{Ber}(0.5)$ process, the GE capacity is determined by a limiting expression, rather than a closed form formula \cite{mushkin1989capacity}.

We apply the proposed scheme to estimate the capacity of the GE channel and compare our results with a consistent estimate of $H(\YY|\XX):=\lim_{n\to\infty}\frac{1}{n}H(Y^n|X^n)$ from a long input-output sequence \cite{rezaeian2005computation}.
The method assumes the elements of $\XX$ are distributed according to the capacity-achieving distribution, i.e., $X_t\sim\mathrm{Ber}(0.5)$.
As seen in Figure \ref{fig:GE}, our method achieves the capacity for a variety of state transition values.

\subsubsection{\texorpdfstring{\underline{Ising Channel}}{Ising Channel}} 
\begin{figure}[t]
  \begin{subfigure}[ht]{.47\textwidth}
    \centerline{\includegraphics[trim={20pt 1pt 20pt 16.4pt}, clip, width=\linewidth]{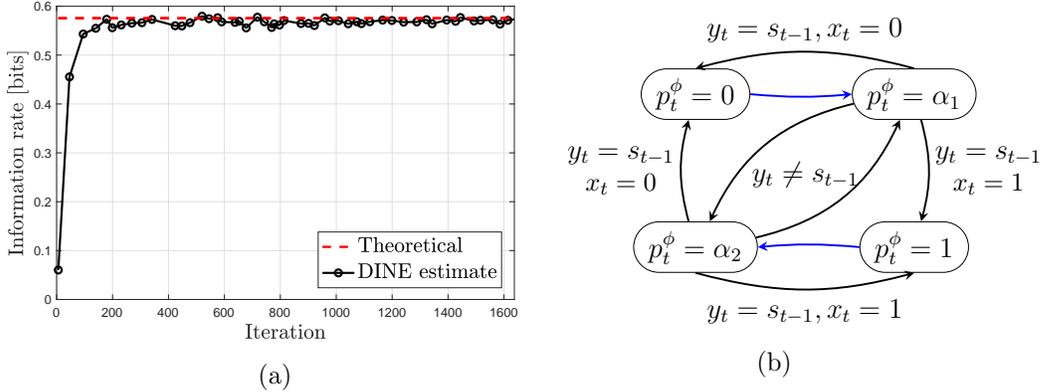}}
    \caption{}
    \label{fig:Ising_convergence}
  \end{subfigure}
  \begin{subfigure}[ht]{.37\textwidth}
    \centering
     \scalebox{0.9}{\begin{tikzpicture}[state/.style={circle, draw, minimum size=2cm}]

\node[rounded rectangle, draw] (1) {\large $p^\phi_t=0$};

\node[rounded rectangle, draw, right=1.5cm of 1] (2) {\large $p^\phi_t=\alpha_1$};

\node[rounded rectangle, draw, below=1.5cm of 1] (3) {\large $p^\phi_t=\alpha_2$};

\node[rounded rectangle, draw, right=1.5cm of 3] (4) {\large $p^\phi_t=1$};

\draw{
(2.north) edge[->,>=stealth,thick,right,bend right=15] node[above = 0mm]{\large $y_t=s_{t-1}, x_t=0$} (1.north)

(1.east) edge[->,>=stealth,thick,right,bend right=5, blue]  (2.west)

(3.south) edge[->,>=stealth,thick,right,bend right=15] node[below = 0mm]{\large $y_t=s_{t-1}, x_t=1$} (4.south)

(4.west) edge[->,>=stealth,thick,right,bend right=5, blue]  (3.east)

(2) edge[->,>=stealth,thick,right,bend right=26] node[below = 0mm]{\large} (3)

(3) edge[->,>=stealth,thick,right,bend right=26] node[above left = 0mm and -3.8mm]{\large $y_t\neq s_{t-1}$} (2)

(3) edge[->,>=stealth,thick,right, bend left=15] node[above left = -5mm and -0.5mm, align=center]{\large $y_t= s_{t-1}$\\$x_t=0$} (1)

(2) edge[->,>=stealth,thick,right, bend left=15] node[above right = -5mm and -0.5mm, align=center]{\large $y_t= s_{t-1}$\\$x_t=1$} (4)
};

\end{tikzpicture}}
    \caption{}
    \label{fig:pmf_model_structure}
  \end{subfigure}
  \caption{Ising Channel. Figure (a) presents the DINE loss convergence on the Ising channel with feedback and Figure (b) presents the learned PMF model for $(\alpha_1,\alpha_2) = (0.456, 0.570)$. A blue arrow denotes a transition that occurs for any value of $(x_t,y_t,s_t)$.}
\end{figure}

The binary Ising channel is a unifilar FSC, that evolves according to\footnote{When the channel model is known, $S_t$ is known at the encoder for any time step since the channel is unifilar.} 
\begin{equation}\label{eq:Ising_eq}
    Y_t = \begin{cases}
      \sZ_{1/2}(X_t), & \text{if } S_{t-1}=0\\
      \sS_{1/2}(X_t),& \text{otherwise}
    \end{cases}, \qquad S_t =  X_{t-1},
\end{equation}
where $\sZ_{1/2}$ and $\sS_{1/2}$ denote the Z- and S-channels with probability $1/2$.
The authors of \cite{elishco2014capacity} compute the capacity of the Ising channel using dynamic programming algorithms over a quantized state space. We estimate the capacity via Algorithm~\ref{alg:di_opt}, which converges to the ground truth capacity after a relatively small number of iterations; cf. Figure \ref{fig:Ising_convergence}. 

We further evaluate our method by analysing the structure of the learned PMF. We construct a long trajectory $(p^{\phi,n}, x^n, s^n, y^n)$
and perform $k$-means clustering of $p^{\phi,n}$ for $k=4$.
We then examine the evolution of clustered $p_t^{\phi}$ according to past $(p^{\phi,t-1}, x^{t-1}, s^{t-1}, y^{t-1})$.
In this case, $p^\phi_t\in[0,1]$ and is treated as the parameter of a Bernoulli distribution.
The structure of $p^\phi_t$ is presented in Figure \ref{fig:pmf_model_structure}.
We note that the joint estimation-optimization procedure results in a nontrivial input PMF evolution, whose structure coincides with the analytical capacity-achieving coding scheme proposed in \cite{elishco2014capacity}.

\subsubsection{\texorpdfstring{\underline{Trapdoor Channel}}{Trapdoor Channel}} 
\begin{figure}[t]
\begin{subfigure}[ht]{0.51\textwidth}
    \centerline{\includegraphics[trim={20pt 1pt 20pt 16.4pt}, clip, width=\linewidth]{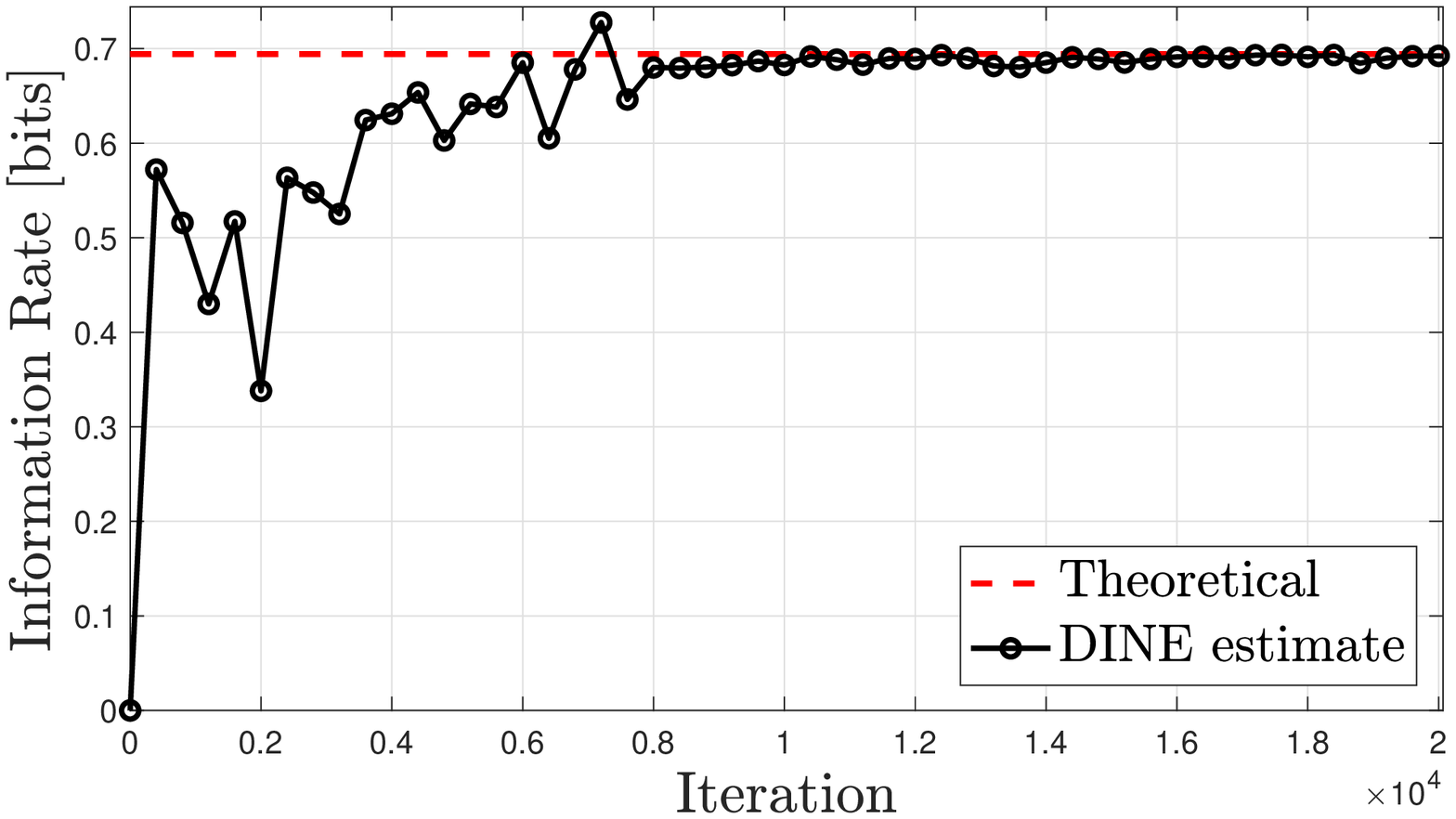}}
    \caption{}
    \label{fig:trapdoor_fb}
  \end{subfigure}
  \begin{subfigure}[ht]{.49\textwidth}
    \centerline{\includegraphics[trim={20pt 1pt 20pt 16.4pt}, clip, width=\linewidth]{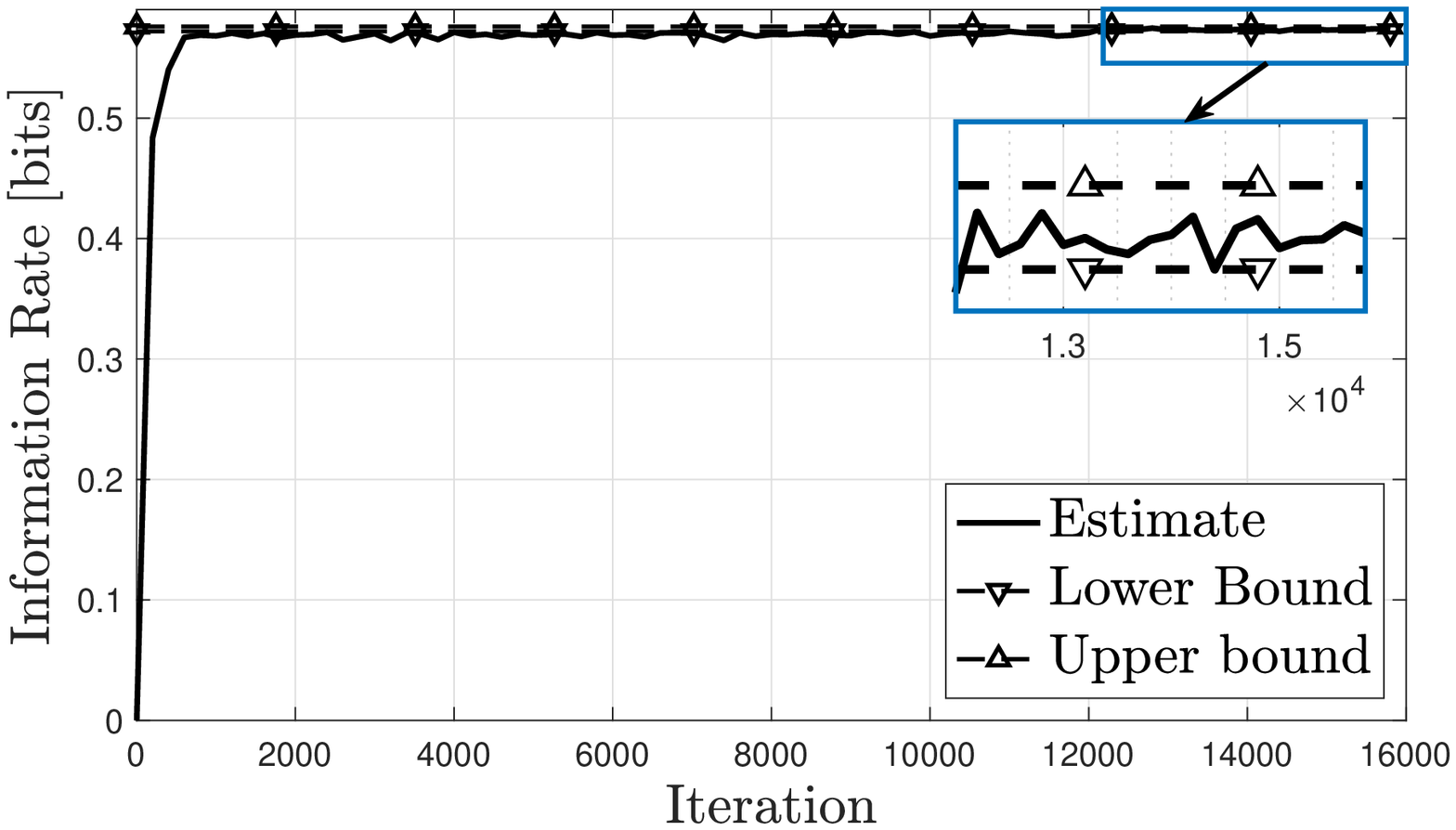}}
    \caption{}
    \label{fig:trapdoor_ff_convergence}
  \end{subfigure}
  \caption{Figures (a) and (b) present DINE loss convergence to the feedback and feedforward trapdoor channel capacities ,respectively.
  Estimated feedforwad capacity is compared with upper and lower bounds from \cite{huleihel2021computable} and \cite{kobayashi2006capacity}, respectively, and the estimated feedback capacity is compared with the analytical solution from \cite{permuter2008capacity}.}
  \label{fig:trapdoor}
\end{figure}

The trapdoor channel is a unifilar FSC whose state evolves according to $S_t = S_{t-1}\oplus X_t \oplus Y_{t}$, where $\oplus$ denotes the binary XOR operation, and its output is given by \eqref{eq:Ising_eq}.
We estimate both the feedforward and feedback capacities of this channel via Algorithm~\ref{alg:di_opt}.
For the feedback capacity, the algorithm converges to the analytic solution from \cite{permuter2008capacity}, as presented in Figure~\ref{fig:trapdoor_fb}.

The feedforward capacity of the trapdoor channel is an open problem. Upper and lower bounds on the capacity value were provided in \cite{huleihel2022capacity}  and \cite{kobayashi2006capacity}; the former used bounds on the delayed feedback capacity, while the latter employ a Blahut-Arimoto-type algorithm, respectively. 
Figure \ref{fig:trapdoor_ff_convergence} shows that the DINE convergence between these known bounds. In particular, this yields a new and improved estimate for the feedforward capacity of the trapdoor channel. Averaging over several runs, we arrive at the value of $\hat{C}^{\mathsf{FF}}_{\text{Trapdoor}} \approx 0.57246$.

\subsubsection{\texorpdfstring{\underline{NOST and POST Channels}}{NOST and POST Channels}} As a last example, we consider the Noisy Output is the STate (NOST) and Previous Output is the STate (POST) channels.
The channel outputs are given by \eqref{eq:Ising_eq}, but with an arbitrary parameter $p\in[0,1]$ (rather than $p=1/2$).
The NOST channel state evolves stochastically according to $S_t= \sZ_\eta(Y_t)$ for $\eta\in[0,1]$, and the POST channel is the special case where $\eta=0$. 

\begin{figure}[t]
  \begin{subfigure}[ht]{.5\textwidth}
    \centerline{\includegraphics[trim={12pt 1pt 20pt 15pt}, clip, width=\linewidth]{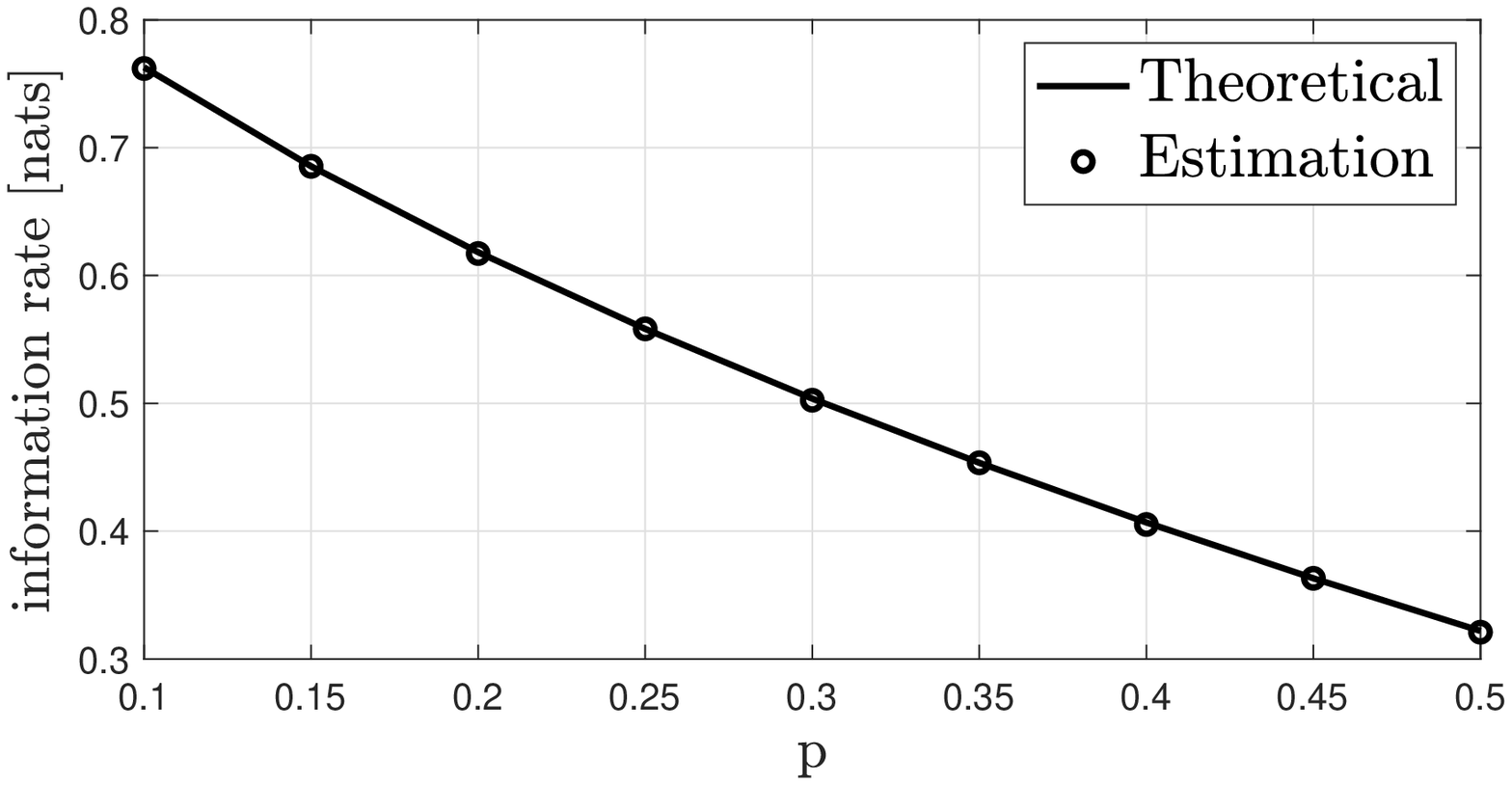}}
    \caption{}
    \label{fig:post}
  \end{subfigure}
  \begin{subfigure}[ht]{0.5\textwidth}
    \centerline{\includegraphics[trim={12pt 1pt 20pt 15pt}, clip, width=\linewidth]{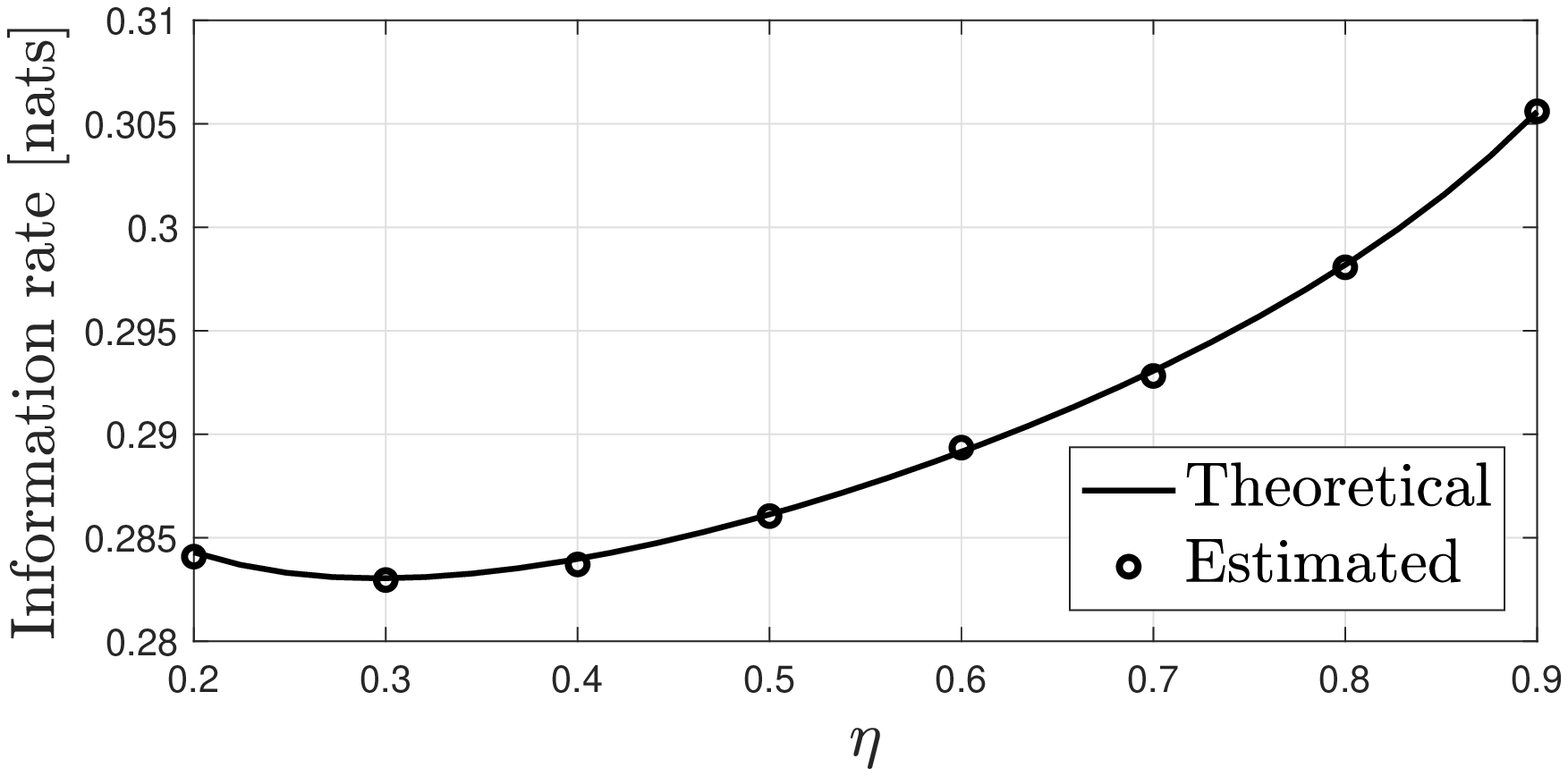}}
    \caption{}
    \label{fig:nost}
  \end{subfigure}
  \caption{Figure (a) presents the POST capacity estimate vs. the channel parameter $p$, compared with the analytical capacity value;
    Figure (b) presents the estimated capacity for the NOST channel versus the flip probability $\eta$, compared with the method in \cite{shemuel2021feedback}.}
  \label{fig:post_nost}
\end{figure}

We use Algorithm~\ref{alg:di_opt} to estimate the feedforward capacity of the POST channel and the feedback capacity of the NOST channel, considering various values of $p$ and $\eta$, respectively.
Figure \ref{fig:post_nost} compares of results to those from \cite{permuter2014capacity} for the POST and \cite{shemuel2021feedback} for the NOST channel.
The authors of \cite{permuter2014capacity} show that both the feedforward and feedback capacities of the $\sf{POST}(p)$ channel are equal to the capacity of the Z channel with the same parameter.
However, the capacity-achieving distribution of the feedforward POST channel can, in general, have infinite memory. This fact together with the accuracy of our capacity estimates testify to the expressiveness of our input distributions model. 

\section{Application 2: Feedback Capacity Bound via Q-Graphs}\label{sec:q_graph}
We develop a method for calculating lower and upper bounds on the feedback capacity of unifilar FSCs, building on the tools from \cite{sabag2016single,sabag2020graph}. The method uses the outputs of Algorithm~\ref{alg:di_opt} to treat an optimization problem involving a structured auxiliary random variable, termed the $Q$-graph.
We begin with an introduction to $Q$-graphs and their utility for calculating capacity bounds. We then argue to the  complexity of the current graph search approach is prohibitive and propose a $Q$-graph approximation algorithm based on the optimized PMF generator from Algorithm~\ref{alg:di_opt}. We demonstrate the performance of our approach both in terms of the estimated $Q$-graphs and the resulting capacity bounds. 
The proposed method couples the capacity estimate with a methodology to calculate upper and lower bounds thereof.

\subsection{\texorpdfstring{Background: $Q$-Graphs}{Background: Q-Graphs}}
For a finite channel output alphabet $\cY$, a $Q$-graph is a directed connected graph with $|\cQ_g|$ nodes and $|\cY|$ distinct outgoing edges from each of the nodes, each uniquely labeled $y\in\cY$.
The node transition on a $Q$-graph is given by a time invariant deterministic function $f_Q:\cQ_g\times\cY\to\cQ_g$.
The authors of \cite{sabag2020graph} showed that for any given $Q$-graph $Q_g$, the feedback capacity of a unifilar FSC can be bounded from both above and below by solving a corresponding optimization problem over some set of input distributions.
Denoting the corresponding bounds by $\underline{\cL}(Q_g)$ and $\overline{\cL}(Q_g)$,
we have \cite[Theorems~2,3]{sabag2016single}: 
\begin{equation}\label{eq:ub_lb_q_graph}
    \underline{\cL}(Q_g)\leq\sC_{\mathsf{FB}}\leq\overline{\cL}(Q_g),
\end{equation}
where both $\underline{\cL}(Q_g)$ and $\overline{\cL}(Q_g)$ are given by $\sup_{P_{X|S,Q}\in\cP}\sI(X,S;Y|Q)$ but with a different set of distributions $\cP$.
For $\overline{\cL}(Q_g)$, we take $\cP=\cP_Q$, which is the set of input distributions that induce a unique stationary joint process $(S_i,Q_i)_{i\in\NN}$. For $\underline{\cL}(Q_g)$, a subset of $\cP_Q$ that admits some Markov criteria is considered.
In practice, given a graph $Q_g$, both upper and lower bounds are obtained by numerically solving the aforementioned problem (see \cite{sabag2020graph} for more details).
Our objective is to find $Q_g$ that either maximizes $\underline{\cL}(Q_g)$ or minimizes $\overline{\cL}(Q_g)$.\footnote{$|\cQ_g|<\infty$ implies that the respective $\argmin$ and $\argmax$ are non-empty, however, they are generally not guaranteed to coincide or overlap.}


\subsection{Existing Method and its Complexity}
 The authors of \cite{sabag2020graph} propose an enumeration-based variant of an exhaustive search to find the best $Q$-graph for a given cardinality $\cQ_g$, termed graph-pooling. 
 %
 %
 This method requires solving both upper and lower bound optimization problems for all considered $Q$-graphs.
As discussed in \cite{sabag2020graph}, bounds on the cardinality of $\cQ_g$ are currently unknown. Consequently, the graph-pooling runtime is, in general, unbounded; if the search over all graphs with $|\cQ_g|$ nodes did not yield tight bounds on the feedback capacity, we continue to search over all graphs of size $|\cQ_g|+1$.
The next lemma shows that the number $Q$-graphs that the graph-pooling approach must consider is exponential in $|\cQ_g|$ (see Section \ref{appendix:GP_complex_bound} for the proof).
\begin{lemma}[Complexity lower bound on graph-pooling method]\label{lemma:GP_complex_bound}
Fix $|\cQ_g|$ and let $N_{\sf{GP}}$ be the number of $Q$-graphs of size $|\cQ_g|$ considered in the graph-pooling method. Then,
$N_{\sf{GP}} \geq e^{|\cQ_g|\log |\cQ_g|}$.
\end{lemma}
Lemma \ref{lemma:GP_complex_bound} suggests that finding a good $Q$-graph with large cardinality is computationally burdensome, even if each optimization problems have relatively low complexity. To overcome this, we next propose an RNN-based approximation method for potentially optimal $Q$-graphs that uses the outputs of Algorithm~\ref{alg:di_opt}, whose training time is independent of $|\cQ_g|$.

\subsection{\texorpdfstring{$Q$-graph Approximation Method}{Q-graph Approximation Method}}
A natural choice for a $Q$-graph is $Q_g=p_{S_t|Y^t}$, as it is the MDP state process in the solution of several unifilar FSCs \cite{permuter2008capacity,elishco2014capacity,sabag2015feedback,shemuel2021feedback}.
For these FSCs it was shown that $p_{S_t|Y^t}$ takes values in a finite subset of $\Delta_{|\cS|}$, which limits the $Q$-graph search space. In situations where such bounds are not available, our method will produce a quantized proxy of $p_{S_t|Y^t}$.
The procedure first approximates $p_{S_t|Y^t}$ using an LSTM network via a supervised learning scheme, and then extracts the graph structure from the learned approximation via $k$-means.

\begin{algorithm}[t]
\caption{$Q$-graph structure and $\CFB$ bounds calculation}
\label{alg:q_est}
\textbf{input:} Discrete channel, optimized PMF generator $\pmf$, and $|\cQ_g|$\\
\textbf{output:} $\hat{C}_{\mathsf{LB}}, \hat{C}_{\mathsf{UB}}$.
\algrule
\begin{algorithmic}
\State Initialize $g_{\psi}$ with parameters $\psi$ and a learning rate $\gamma$.
\State \textbf{Step 1: Train $\bm{g_{\psi}}$}
\Repeat
\State Compute $(S^n,Y^n)$ using $\pmf$ and channel.
\State Compute $\{q^\psi_t\}_{t=1}^n$ using $g_{\psi}$
\State Compute CE loss \eqref{eq:ce_loss} and update $\psi$: \\
\hspace{\algorithmicindent}\hspace{\algorithmicindent}$\psi \leftarrow \psi - \gamma\nabla_{\psi}\cL_{\mathsf{CE}}(Y^n,S^n,\psi)$ 
\Until{convergence}
\State \textbf{Step 2: Obtain bounds}
\State Generate a sequence $(q^{\psi,n},Y^n)$
\State Perform $k$-means clustering on $q^{\psi,n}$ 
\State Compute $M_Q$ from $(q^{\psi,n},Y^n)$ 
\State Compute $\hat{C}_{\mathsf{LB}}, \hat{C}_{\mathsf{UB}}$ from $M_Q$\\
\Return $\hat{C}_{\mathsf{LB}}, \hat{C}_{\mathsf{UB}}, M_Q$
\end{algorithmic}
\end{algorithm}
 
\subsubsection{\texorpdfstring{\underline{Mapping Approximation}}{Mapping Approximation}} We devise an RNN-based generative model $g_\psi:\Delta_{|\cS|}\times\cY\to \Delta_{|\cS|}$ with parameters $\psi\in\RR^d$.
The model receives a sequence of channel outputs $Y^n$, generated by the optimized PMF model and the channel, and recursively calculates a sequence $(q^\psi_t)_{t=1}^n$.
The model is initialized with $q_0=0$, and evolves according to
\[
    q^\psi_t = g_{\psi}(q^\psi_{t-1},Y_t), \quad t=1,\dots,n.
\]
The model is trained to approximate the evolution of $p_{S_t|Y^t}$ by minimizing the \textit{categorical cross entropy}, given by
\begin{equation}\label{eq:ce_loss}
    \cL_{\mathsf{CE}}(Y^n,S^n,\psi) :=
    -\sum_{t=1}^n e_{S_t}^{\sT}\log q^\psi_t
    = -\sum_{t=1}^n \log q^\psi_{t,S_t},
\end{equation}
where $e_{S_t}$ is a one-hot encoding of $S_t$, and $q^\psi_{t,S_t}$ is $S_t$-th coordinate of $q^\psi_t$.

\subsubsection{\texorpdfstring{\underline{Trajectory Analysis}}{Trajectory Analysis}}
Having an RNN approximation of $p_{S_t|Y^t}$, we aim to extract the underlying graph structure. 
To that end, we construct a long sequence $(q^{\psi,n},Y^n)$ and apply $k$-means clustering\cite{jain1988algorithms} to $q^{\psi,n}$ such that each cluster represents an approximated graph node.
Next, we label the edges by taking the most frequent transition between each two nodes\footnote{A generalized version of $Q$-graphs can consider randomized mappings, therefore also including less frequent transitions.}. 
The graph transition information is stored in a $3$-dimensional binary adjacency data structure, denoted $M_Q$, such that $M_Q(i,j,k)=1$ if the edge from node $i$ to node $j$ exists with label $y=k$.

We plug the $Q$-graph corresponding to $M_Q$ into the optimization problems from \cite{sabag2020graph} and compute $\underline{\cL}(M_Q)$ and $\overline{\cL}(M_Q)$.
When the resulting bounds are not tight we can repeat the analysis step for larger values of graph nodes $k$, using the same trajectory  $(q^{\psi,n},Y^n)$.
The complete scheme is presented in Algorithm \ref{alg:q_est}.
We stress that while Algorithm~\ref{alg:di_opt} does not require access to the channel state sequence, the proposed bounds calculation method does rely on this information.



\subsection{Empirical Results}
\begin{figure}[t]
    \centering
    \hspace{-1cm}
    \begin{subfigure}[b]{0.33\textwidth}
        \centering
        \scalebox{.75}{\begin{tikzpicture}[state/.style={circle, draw, minimum size=2cm}]
\node[rounded rectangle, draw] (1) {$\hat{Q}_{g,0}=0.224$};
\node[rounded rectangle, draw, below =2cm  of 1, align=left] (2) {$\hat{Q}_{g,1} = 0.398$};
\node[rounded rectangle, draw, right=2cm of 1] (4) {$\hat{Q}_{g,3}=0.781$};
\node[rounded rectangle, draw, below =2cm of 4, align=left] (3) {$\hat{Q}_{g,2}=0.606$};
\draw{
(1) edge[->,>=stealth,thick,loop left,left=1] node{$1$} (1)
(2) edge[->,>=stealth,thick,left,bend left=10] node{$1$} (1)
(3) edge[->,>=stealth,thick,left,bend right=10,near start] node{$1$} (1)
(4) edge[->,>=stealth,thick,loop right,right=1] node{$0$} (4)
(3) edge[->,>=stealth,thick,left,bend right=10] node{$0$} (4)
(2) edge[->,>=stealth,thick,left,bend left=10,near start] node{$0$} (4)

(1) edge[->,>=stealth,thick,right,bend right=10,near start] node{$0$} (3)
(4) edge[->,>=stealth,thick,right,bend left=10,near start] node{$1$} (2)
};
\end{tikzpicture}}
        \caption{Estimated $Q$-graph.}
        \label{fig:q_trapdoor}
    \end{subfigure}
    \hspace{3cm}
    \begin{subfigure}[b]{0.33\textwidth}
        \centering
        \scalebox{.75}{\begin{tikzpicture}[state/.style={circle, draw, minimum size=2cm}]
\node[rounded rectangle, draw] (1) {$p_0=0.236$};
\node[rounded rectangle, draw, below =2cm  of 1, align=left] (2) {$p_1 = 0.382$};
\node[rounded rectangle, draw, right=2cm of 1] (4) {$p_3=0.764$};
\node[rounded rectangle, draw, below =2cm of 4, align=left] (3) {$p_2=0.613$};
\draw{
(1) edge[->,>=stealth,thick,loop left,left=1] node{$1$} (1)
(2) edge[->,>=stealth,thick,left,bend left=10] node{$1$} (1)
(3) edge[->,>=stealth,thick,left,bend right=10,near start] node{$1$} (1)
(4) edge[->,>=stealth,thick,loop right,right=1] node{$0$} (4)
(3) edge[->,>=stealth,thick,left,bend right=10] node{$0$} (4)
(2) edge[->,>=stealth,thick,left,bend left=10,near start] node{$0$} (4)

(1) edge[->,>=stealth,thick,right,bend right=10,near start] node{$0$} (3)
(4) edge[->,>=stealth,thick,right,bend left=10,near start] node{$1$} (2)
};
\end{tikzpicture}}
        \caption{Analytical MDP state. }
    \label{fig:dp_trapdoor}
    \end{subfigure}
    
    \caption{Trapdoor channel. Comparison between the estimated structure (a) and the dynamic program optimized MDP state transition from \cite{permuter2008capacity} (b).}
    \label{fig:trapdoor_q_compare}
\end{figure}
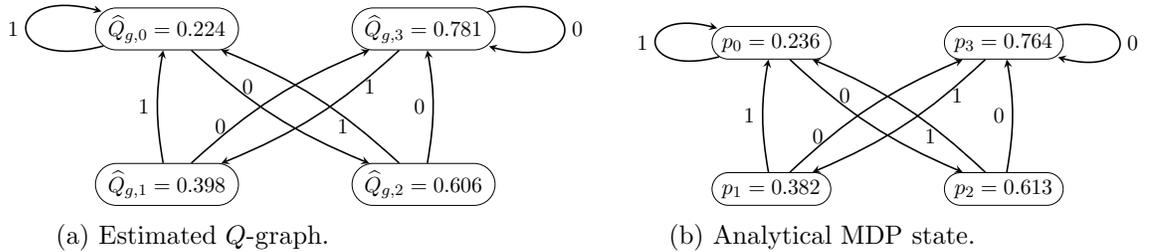

We demonstrate the performance of Algorithm \ref{alg:q_est} on the Trapdoor and Ising channels, as the structure of $p_{S_t|Y^t}$ under the optimal input distribution is known for these examples.
Figures \ref{fig:Ising_q_comapre} and \ref{fig:trapdoor_q_compare} compare the learned $Q$-graphs with the optimal structure derived in \cite{elishco2014capacity} and \cite{permuter2008capacity}.
The estimated $p_{S_t|Y^t}$ coincides with the one obtained from the analytical solution, although the numerical values associated with the nodes are slightly different.
Nevertheless, these tweaked values do not affect the bound computation~\cite{sabag2020graph}, 
and using Equations (13) and (16) from \cite{sabag2020graph} we obtain the following capacity upper and lower bounds:
\begin{equation*}
    \text{Trapdoor:}\quad \hat{C}_{\mathsf{UB}}-\hat{C}_{\mathsf{LB}} \approx 3.7615\cdot 10^{-08},\quad 
    \text{Ising:}\quad \hat{C}_{\mathsf{UB}}-\hat{C}_{\mathsf{LB}} \approx 2.4334\cdot 10^{-07}
\end{equation*}
The calculated bounds are extremely close, testifying to the potency of the proposed method and providing another useful byproduct of Algorithm~\ref{alg:di_opt}.

\begin{figure}[t]
    \centering
    \begin{subfigure}[b]{0.4\textwidth}
        \centering
        \scalebox{.75}{\begin{tikzpicture}[state/.style={circle, draw, minimum size=2cm}]
\node[rounded rectangle, draw] (1) {$\hat{Q}_{g,0}=0.0012$};
\node[rounded rectangle,draw, below right =2cm and 0.35cm of 1, align=left] (2) {$\hspace{0.8cm}(\hat{Q}_{g,1},\hat{Q}_{g,2})$ \\ $=(0.3732,0.6041)$};
\node[rounded rectangle, draw, right=2cm of 1] (3) {$\hat{Q}_{g,3}=0.9986$};
\draw{
(3) edge[->,>=stealth,thick,above,bend right=10] node{$1$} (1)
(1) edge[->,>=stealth,thick,below,bend right=10] node{$0$} (3)
(1) edge[->,>=stealth,thick,left,bend right=10] node{$1$} (2)
(2) edge[->,>=stealth,thick,right,bend right=10] node{$1$} (1)
(3) edge[->,>=stealth,thick,left,bend right=10] node{$0$} (2)
(2) edge[->,>=stealth,thick,right,bend right=10] node{$0$} (3)

};
\end{tikzpicture}}
        \caption{Estimated $Q$-graph.}
        \label{fig:q_Ising}
    \end{subfigure}
    \hspace{1cm}
    \begin{subfigure}[b]{0.4\textwidth}
        \centering
        \scalebox{.75}{\begin{tikzpicture}[state/.style={circle, draw, minimum size=2cm}]
\node[rounded rectangle, draw] (1) {$p_0=0$};
\node[rounded rectangle, draw,below right =2cm and 0.35cm of 1, align=left] (2) {$\hspace{0.8cm}(p_1,p_2)$ \\ $=(0.379,0.621)$};
\node[rounded rectangle, draw, right=2cm of 1] (3) {$p_3=1$};
\draw{
(3) edge[->,>=stealth,thick,above,bend right=10] node{$1$} (1)
(1) edge[->,>=stealth,thick,below,bend right=10] node{$0$} (3)
(1) edge[->,>=stealth,thick,left,bend right=10] node{$1$} (2)
(2) edge[->,>=stealth,thick,right,bend right=10] node{$1$} (1)
(3) edge[->,>=stealth,thick,left,bend right=10] node{$0$} (2)
(2) edge[->,>=stealth,thick,right,bend right=10] node{$0$} (3)

};
\end{tikzpicture}
        }
        \caption{Analytical MDP state.}
    \label{fig:dp_Ising}
    \end{subfigure}
    
    \caption{Ising channel. Comparison between the estimated structure (a) and the dynamic program optimized MDP state transition from \cite{elishco2014capacity} (b).}
    \label{fig:Ising_q_comapre}
\end{figure}
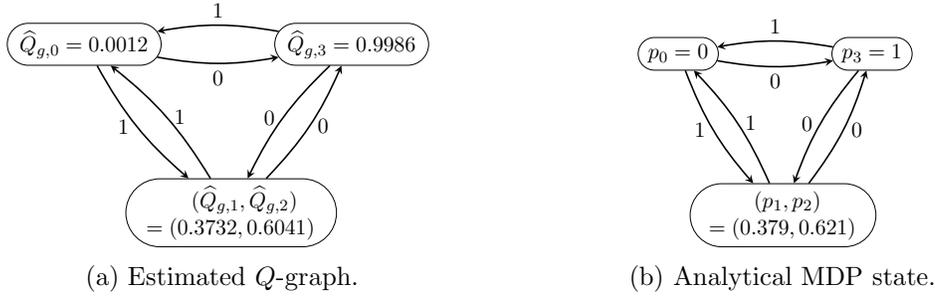

\begin{figure}[t]
    \centering
    \begin{subfigure}[b]{0.43\textwidth}
        \centering
        \includegraphics[ width=\linewidth]{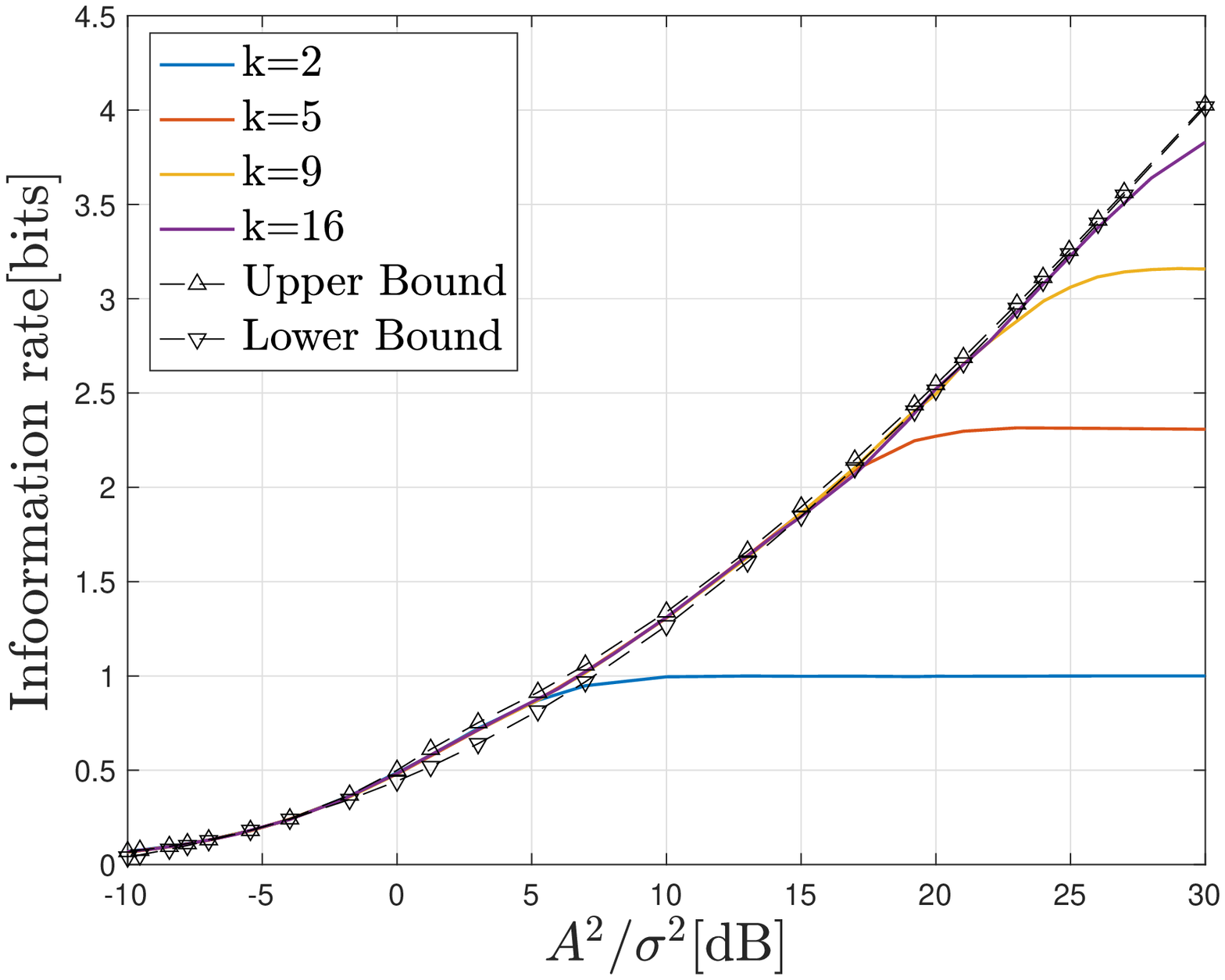}
        \caption{}
        \label{fig:pam_vs_bounds}
    \end{subfigure}
    \begin{subfigure}[b]{0.43\textwidth}
        \centering
        \includegraphics[width=1\textwidth]{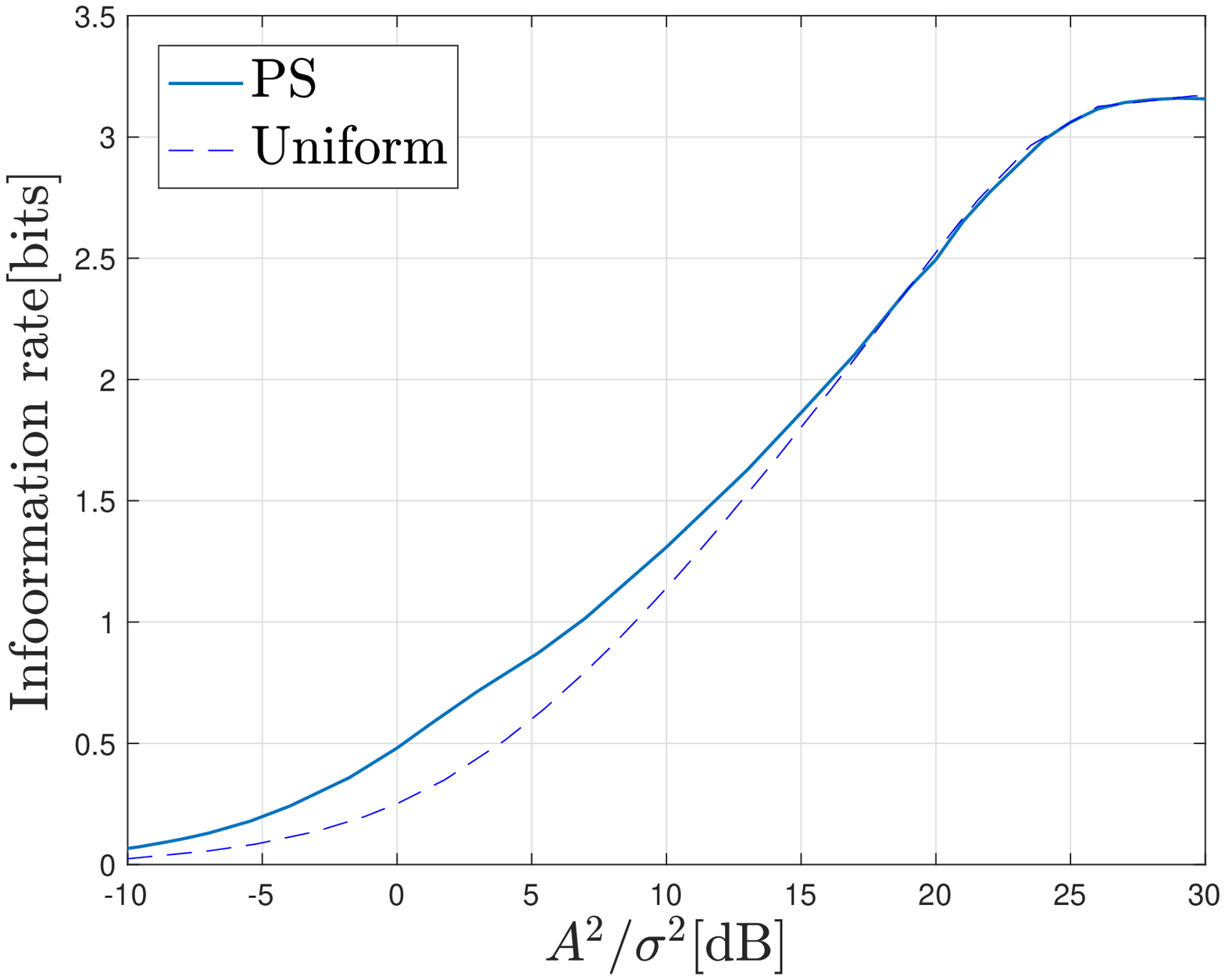}
        \caption{}
        \label{fig:pam_vs_uniform}
    \end{subfigure}
    \caption{Estimated MI for the real valued PP-AWGN. Figure (a) shows a comparison of the optimized MI with capacity upper and lower bounds; and Figure (b) presents a comparison with the MI induced by the uniform distribution for $k=9$.}
    \label{.}
\end{figure}

\section{Application III: Probabilistic Shaping of Constellations}\label{sec:prob_shape}

Conventional communication schemes consider equiprobable constellations \cite[Section~7]{proakis1994communication}.
By applying Algorithm~\ref{alg:di_opt} for MI optimization (see Section \ref{sec:mine_opt}), we propose a non-uniform shaping scheme and show that it outperforms the equiprobable constellation in terms of the communication rate.
Consider the PP-AWGN, given by
\begin{align*}
    &Y = X+Z, \qquad \text{s.t.} \quad|X|\leq A,\quad P_X - a.s.,
\end{align*}
where $A>0$ and $Z$ is a centered Gaussian noise with variance $\sigma^2$. Our goal is to design a discrete constellation for $X$ that maximizes the transmission rate.


\begin{figure}[b]
    \centering
    \begin{subfigure}[b]{0.3\textwidth}
        \centering
        \includegraphics[width=1\textwidth]{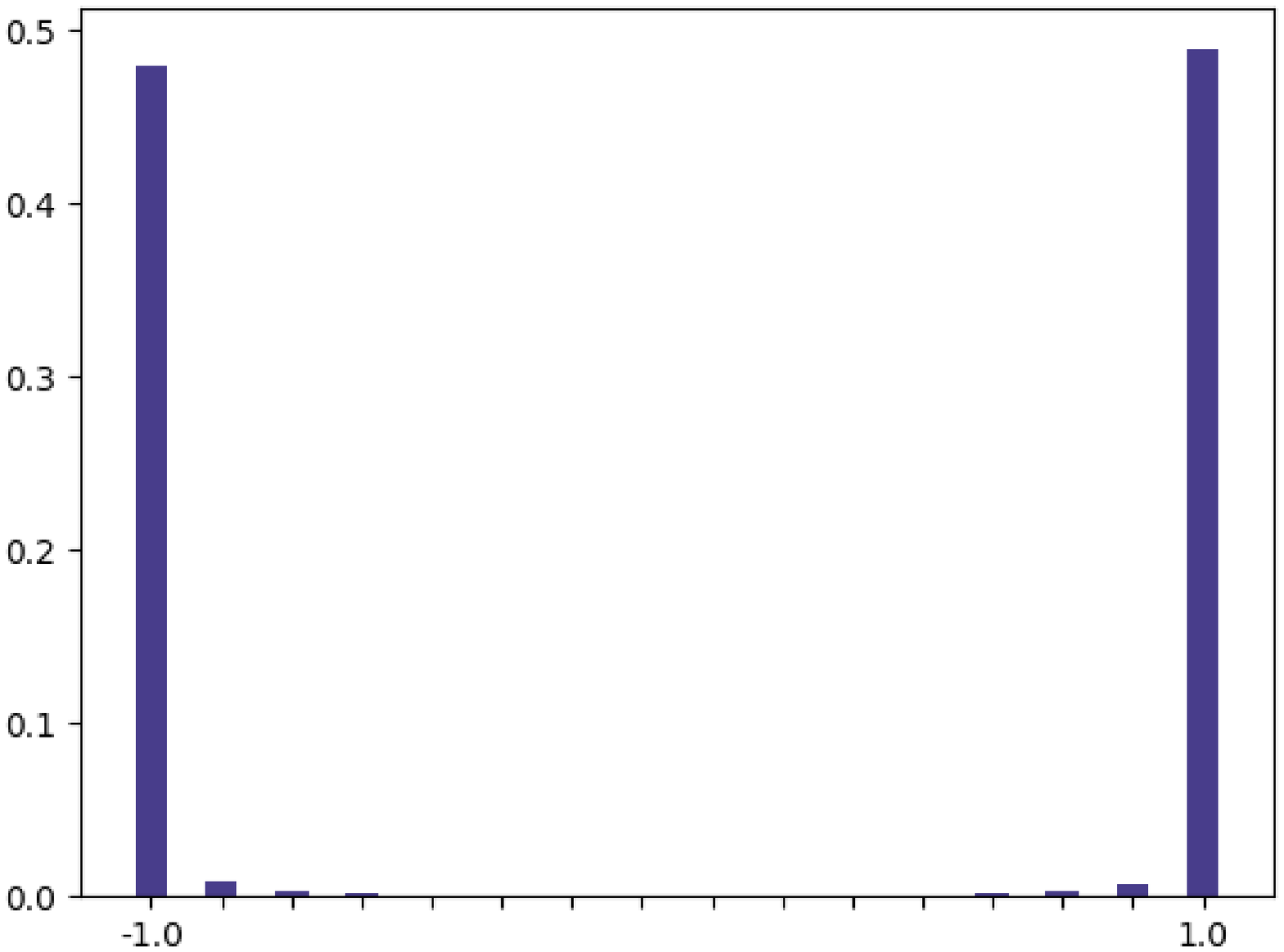}
        \caption{$\pSNR=-10 [dB]$.}
    \end{subfigure}
    \begin{subfigure}[b]{0.3\textwidth}
        \centering
        \includegraphics[ width=\linewidth]{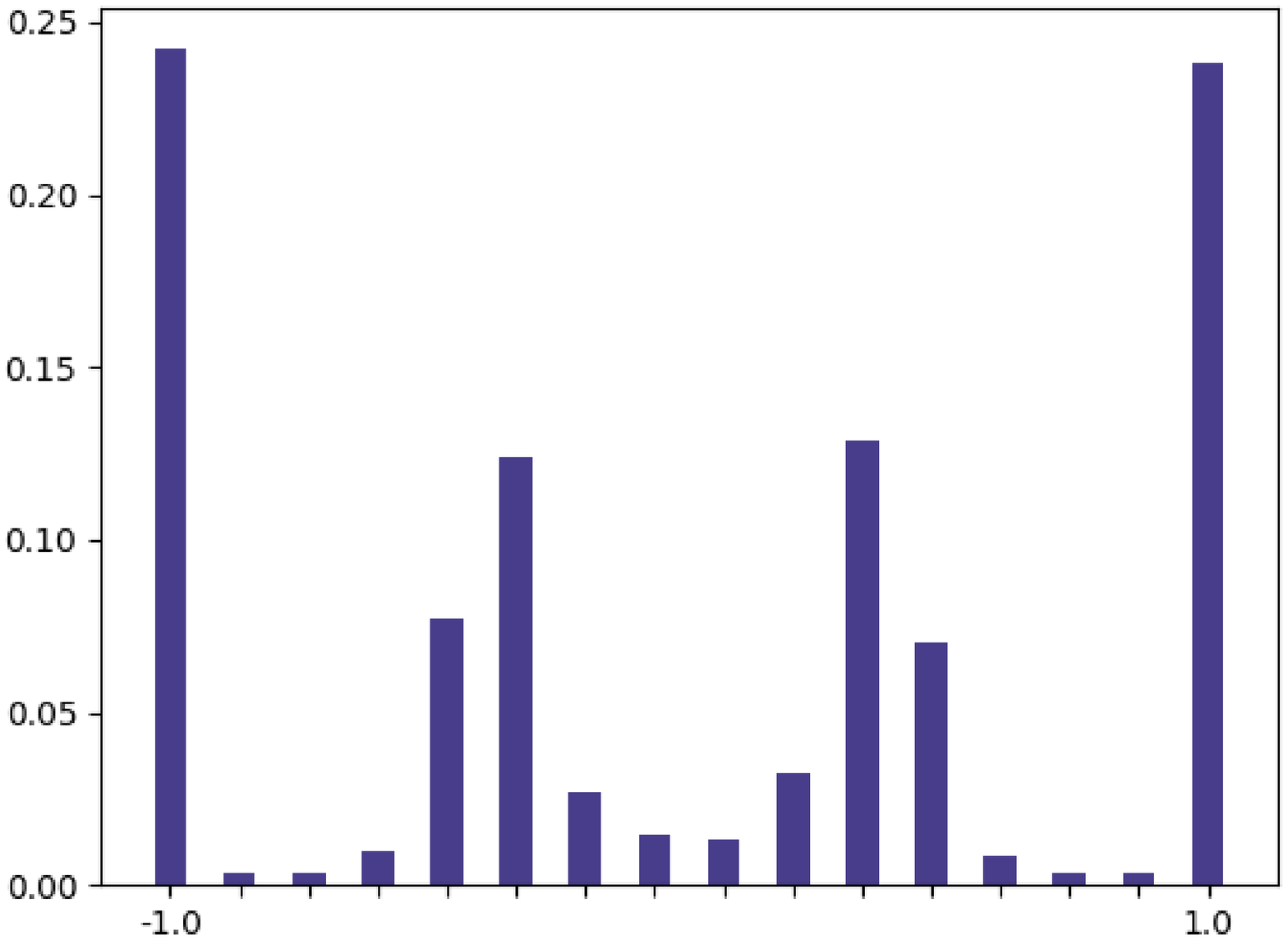}
        \caption{$\pSNR=15 [dB]$}
    \end{subfigure}
    \begin{subfigure}[b]{0.3\textwidth}
        \centering
        \includegraphics[width=1\textwidth]{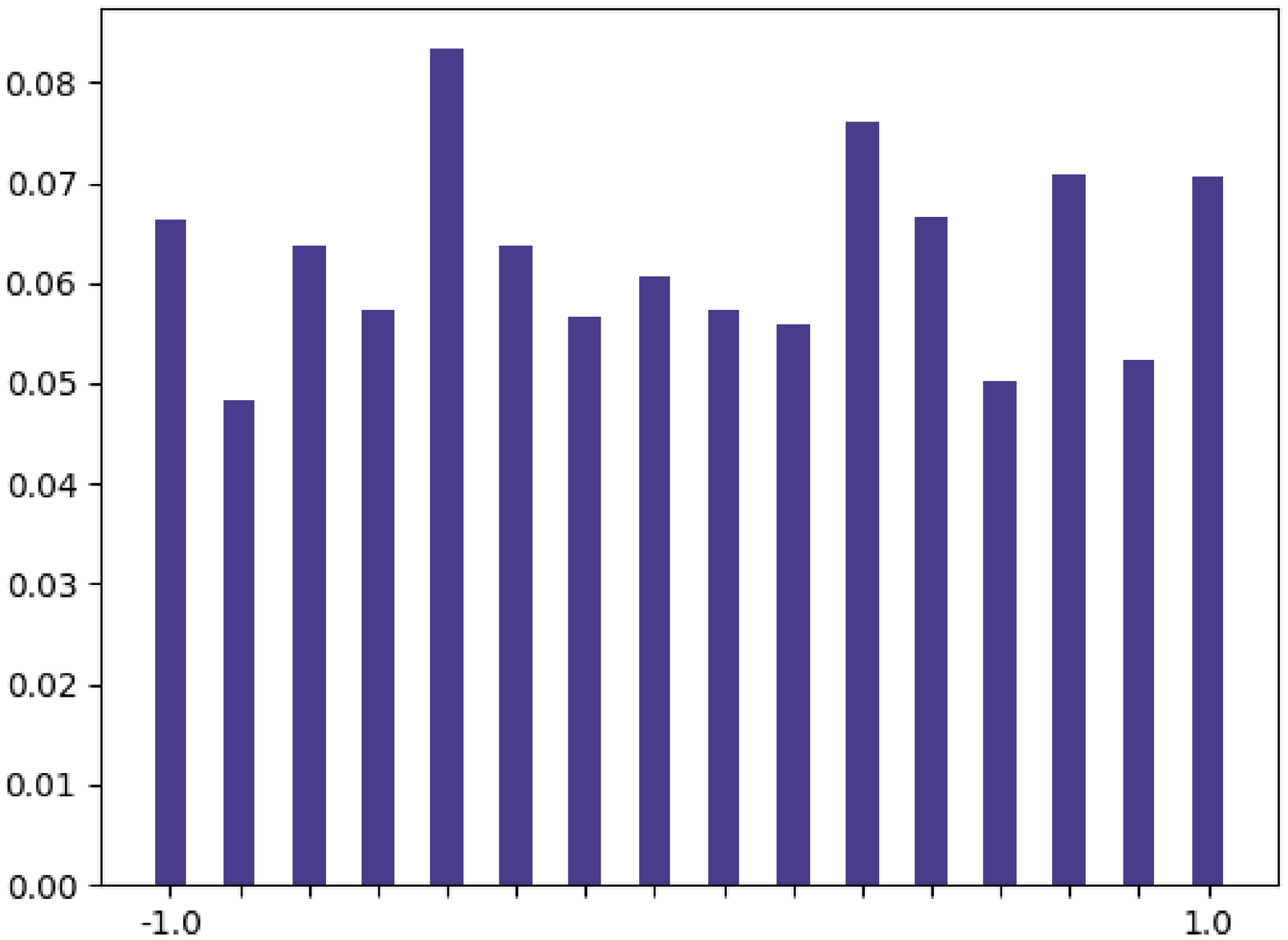}
        \caption{$\pSNR=30 [dB]$}
    \end{subfigure}
    \caption{Learned probabilistic shaping for order $k=16$ and several values of $\pSNR$.}
    \label{fig:PAM_constellations}
\end{figure}

\subsection{Real-Valued AWGN}\label{subsec:real_AWGN}
Smith showed that the capacity of the real-valued PP-AWGN channel is achieved by a discrete distribution supported inside $[-A,A]$, whose cardinality grows with $\pSNR$ \cite{smith1971information}.
We thus consider PAM constellations of different orders within $[-A,A]$.

Figure \ref{fig:pam_vs_bounds} compares the estimated MI to analytical upper \cite{ozarow1990capacity} and lower bounds \cite{thangaraj2017capacity}, for a range of $\pSNR$ values.
Evidently, the MI estimate converges between the bounds. 
Recall that MINE itself serves as a lower bounds the ground truth MI \cite[Remark~5]{tsur2022neural}, posing our estimate as a new numerical lower bound on the PP-AWGN capacity.
We observe that the information rate saturates for each considered constellation orders, as SNR grows; this stems from the source entropy upper bound $\sI(X;Y)\leq H(X)$.
Figure \ref{fig:pam_vs_bounds} therefore reveals the values of $\pSNR$ beyond which a certain order $m$ is no longer optimal.
Figure \ref{fig:pam_vs_uniform} shows a comparison of the estimated optimized MI with the one induced by a uniform distribution over the constellation elements.
It is clear that the learned probabilistic shaping results in higher MI for every considered SNR~value.


The learned probabilistic shaping also corresponds to asymptotic values of the optimal input distribution derived in \cite{smith1971information}. Namely, it was shown that the optimal distribution converges towards a Bernoulli distribution on $\{-A,A\}$ as $\pSNR\to 0$, while for  $\pSNR\to \infty$ the PMF becomes uniform distribution over the entire interval $[-A,A]$. Figure \ref{fig:PAM_constellations} depicts the learned probabilistic shaping for $k=16$, which indeed adheres to the mentioned asymptotic behavior. The intermediate point $\pSNR=12 [dB]$ is an example of when an interpolation between the two extreme cases is used.

\subsection{Complex AWGN}\label{subsec:complex_swgn}

To code for the complex AWGN channel, we consider a rectangular QAM constellations. The peak constraint now becomes as box constraint, whereby $\mathsf{Re}(X)$ and $\mathsf{Im}(X)$ are bound to the one-dimensional peak-power constraint.
Under the box constraint, it is shown in \cite{ikeda2010capacity} that the optimal input $X$ has $\mathsf{Re}(X)$ and $\mathsf{Im}(X)$ independent, due to the independence of the noise components.
Using this fact, we have the capacity bounds
\begin{equation}\label{eq:complex_awgn_bounds}
    2C_{\mathsf{LB}}^{\mathsf{real}} \leq C \leq 2C_{\mathsf{UB}}^{\mathsf{real}},
\end{equation}
where $C_{\mathsf{LB}}^{\mathsf{real}}$ and $C_{\mathsf{UB}}^{\mathsf{real}}$ are the real-valued PP-AWGN capacity bounds from Section \ref{subsec:real_AWGN}.

\begin{figure}
    \centering
    \begin{subfigure}[t]{0.46\textwidth}
        \centering
        \includegraphics[width=1\linewidth]{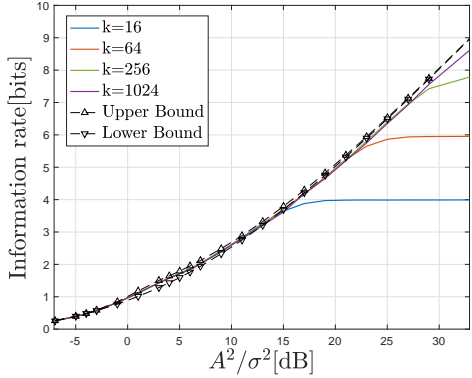}
        \caption{Information rate comparison.}
        \label{fig:qam_vs_bounds}
    \end{subfigure}
    \begin{subfigure}[t]{0.53\textwidth}
        \centering
    \includegraphics[trim={1pt 55pt 1pt 1pt}, clip, width=\linewidth]{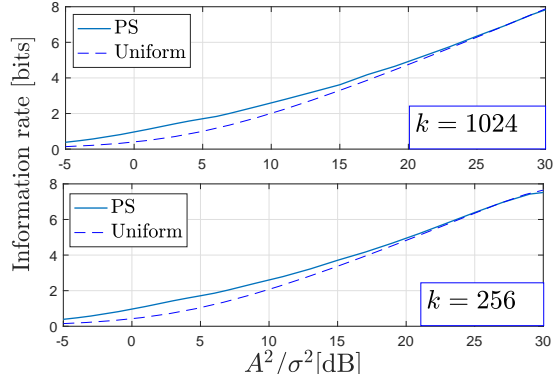}
        \caption{Comparison with uniform.}
        \label{fig:qam_vs_unif1}
    \end{subfigure}
    \label{fig:qam_vs_unif_tot}
    \caption{QAM probabilistic shaping performance. Figure (a) shows a comparison of the optimized MI for several constellation orders with analytical upper and lower bounds from \cite{ozarow1990capacity} and \cite{thangaraj2017capacity}, respectively; Figure (b) shows a comparison of the optimized MI with the MI induced by the uniform distribution.}
\end{figure}

\begin{figure}[t]
    \centering
    \begin{subfigure}[b]{0.3\textwidth}
        \centering
        \includegraphics[width=1\textwidth]{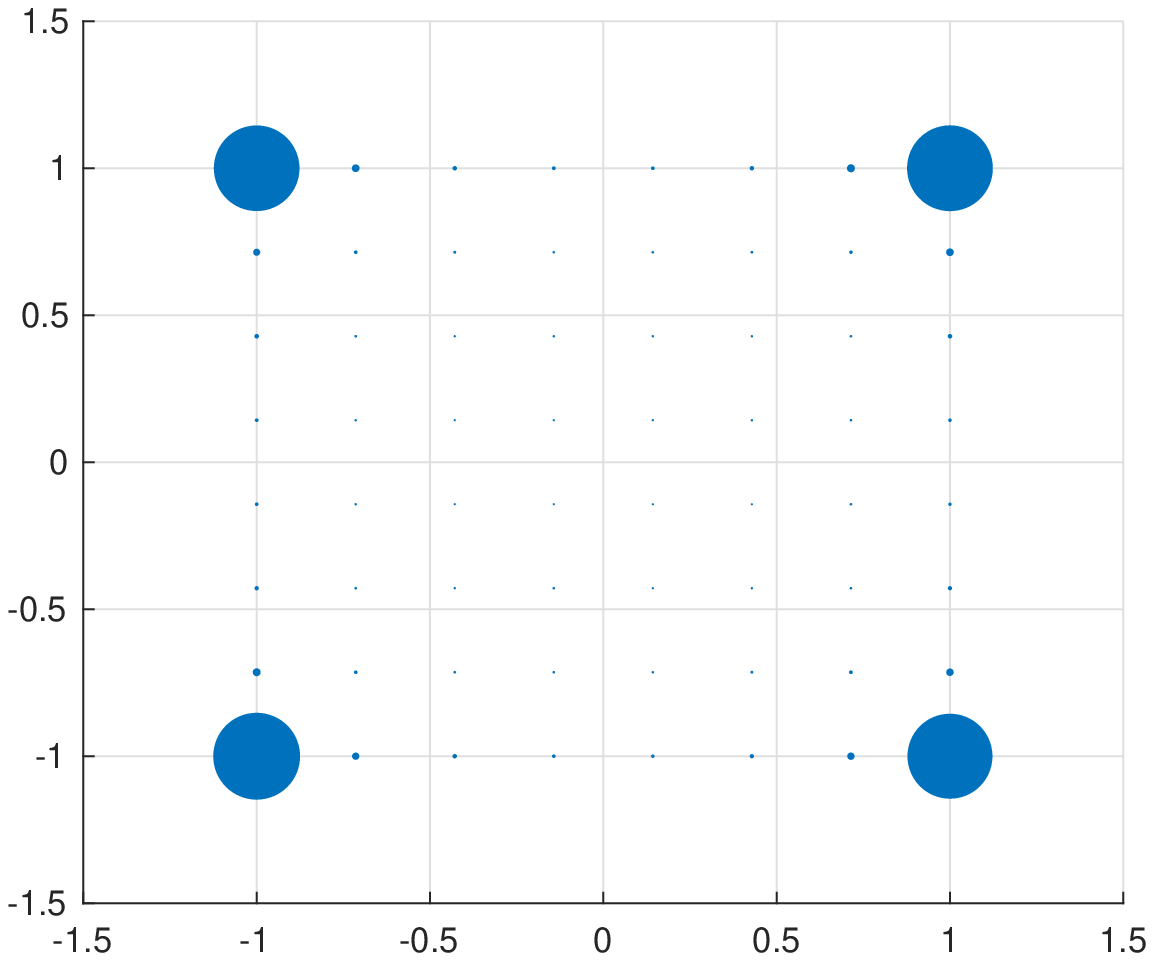}
        \caption{$\pSNR=-10 [dB]$.}
    \end{subfigure}
    \begin{subfigure}[b]{0.3\textwidth}
        \centering
        \includegraphics[ width=\linewidth]{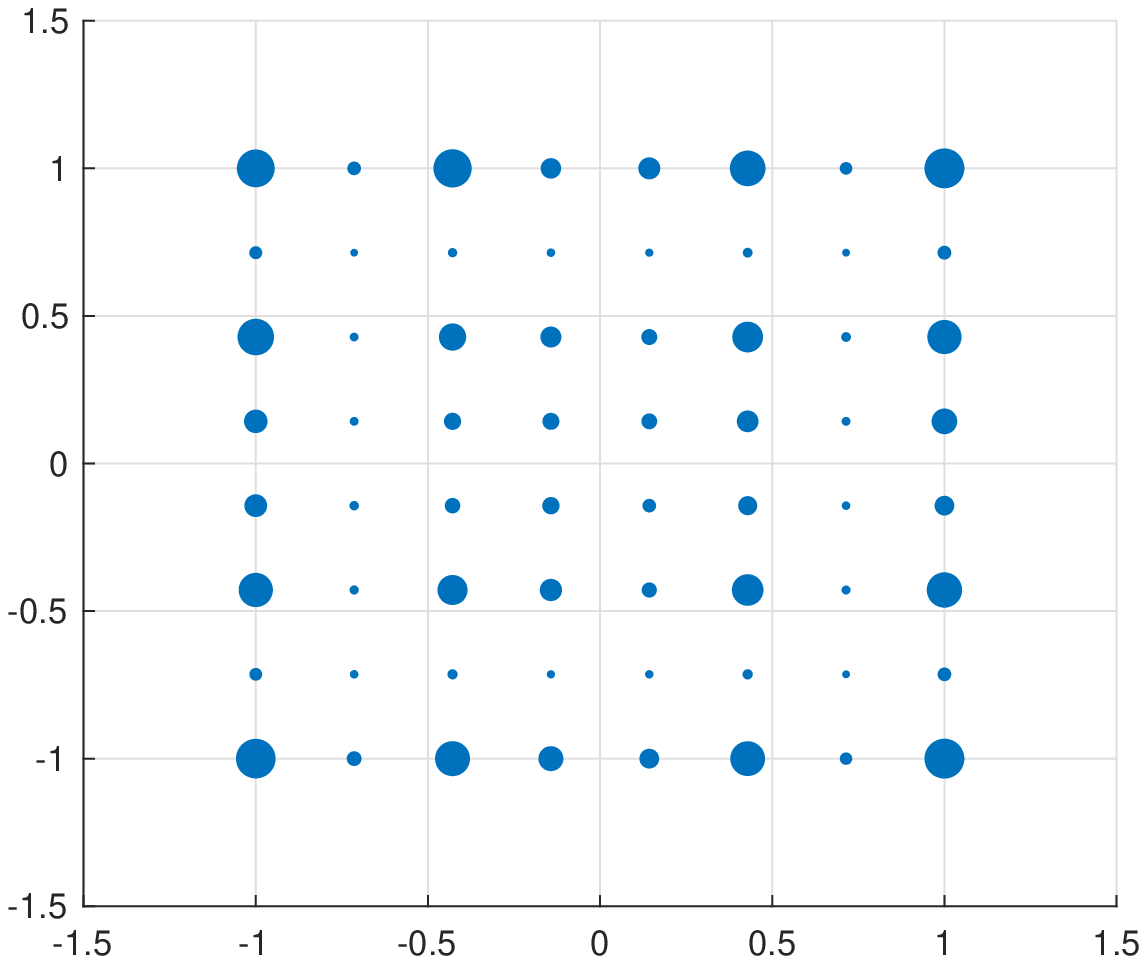}
        \caption{$\pSNR=12 [dB]$}
    \end{subfigure}
    \begin{subfigure}[b]{0.3\textwidth}
        \centering
        \includegraphics[width=1\textwidth]{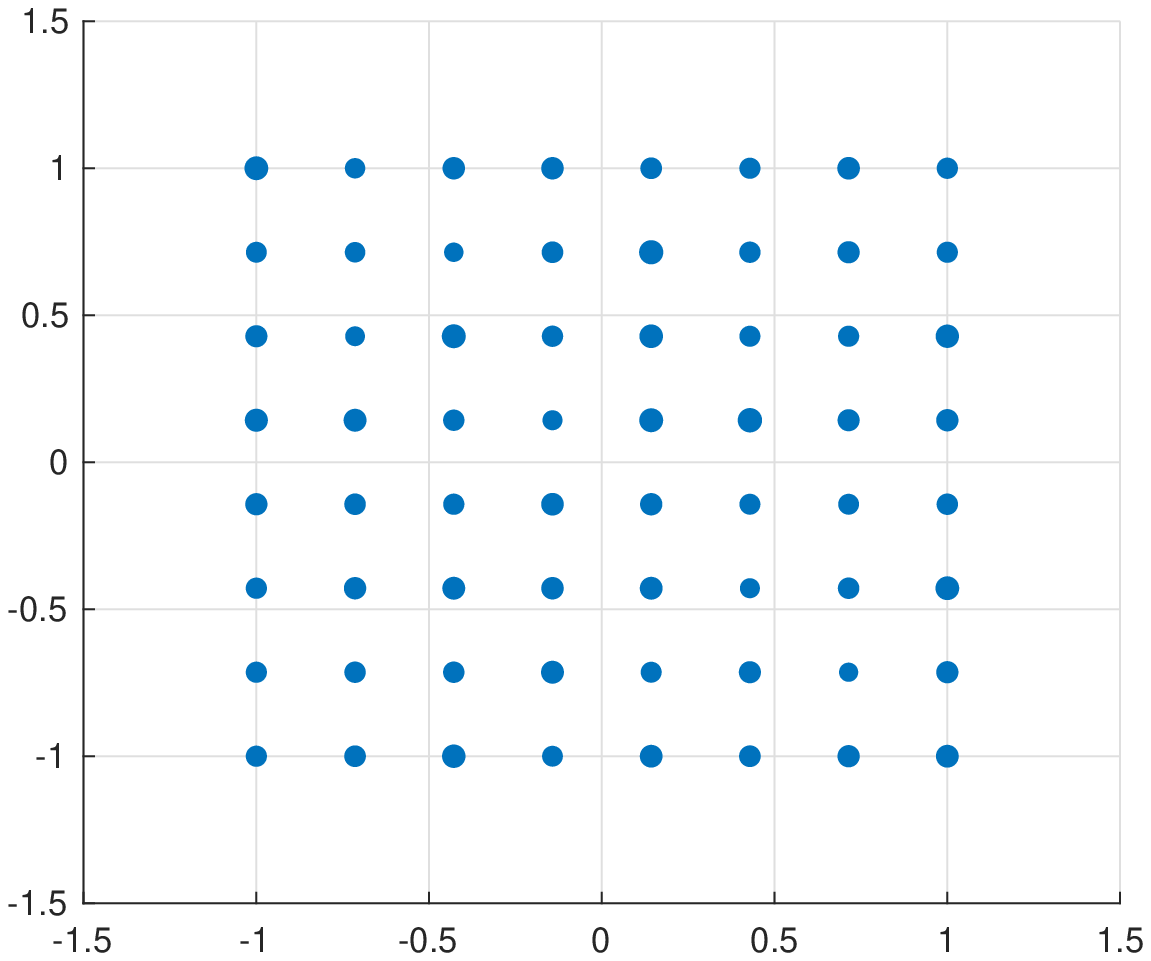}
        \caption{$\pSNR=30 [dB]$}
    \end{subfigure}
    \caption{Learned QAM distribution for several values of $\pSNR$ and $k=64$. The marker size denotes the assigned probability.}
    \label{fig:QAM_constellations}
\end{figure}

Figure \ref{fig:qam_vs_bounds} compares our estimated MI values with the above bounds for several QAM orders.
In Figure \ref{fig:qam_vs_unif1} we compare the MINE estimate with the MI induced by the uniform distribution, calculated via the Gauss-Hermite integral approximation.
It is clear that the learned distribution outperforms the uniform one for all considered $\pSNR$ values.
Finally, Figure \ref{fig:QAM_constellations} shows the learned probabilistic shaping for $k=64$ and several values of $\pSNR$.
Since $\mathsf{Re}(X)$ and $\mathsf{Im}(X)$ are independent, we look for features similar to those observed for the real-valued AWGN channel.
Indeed, we see that for low values of $\pSNR$ the learned probabilistic shaping is uniform on the constellation edges, while as $\pSNR$ grows, the probabilistic shaping shifts towards a uniform distribution over the entire constellation.
Nontrivial input distributions, as shown in \ref{fig:QAM_constellations}(b), are observed for intermediate values of $\pSNR$.

\section{Proofs}\label{sec:proofs}
\subsection{Proof of Theorem \ref{theorem:MDP_formulation}}\label{proof:mdp_formulation}
The proof shows that (i) $Z_t$ evolves as a function of $(Z_{t-1},U_{t-1},W_{t-1})$, (ii) $P_{W_t|W^{t-1},U^{t},Z^{t}} = P_{W_t|W_{t-1},U_{t},Z^{t}}$, and (iii) $\rho(\pi_\phi) = \sI_\phi(\XX\to\YY)$. 
First, the functional relation $Z_t = f(Z_{t-1},U_t,W_t)$ follows by defining $f$ as a concatenation of $Z_{t-1}$ with $(U_t,W_t)$.
For (ii), we have
\begin{align*}
    P_W(W_t=w|W^{t-1}=w^{t-1},Z^{t-1}=z^{t-1},U^{t-1}=u^{t-1}) &= \PP(Y_0=y|X^{0}_{-t}=x^{0}_{-t},Y^{0}_{-t}=y^{0}_{-t})\\
    &=P_W(W_t=w|Z_{t-1}=z_{t-1},U_{t-1}=u_{t-1}).
\end{align*}
To establish (iii), observe that
\begin{align*}
    \rho(\pi_\phi) 
    &=\lim_{N\to\infty}\frac{1}{N}\sum_{t=1}^N \EE\left[\EE\left[\log\frac{P_{Y_0|Y^{-1}_{-t},X^{0}_{-t}}}{P_{Y_0|Y^{-1}_{-t}}}\Big| X^{0}_{-t},Y^{-1}_{-t}\right]\right]\\
    &= \lim_{N\to\infty}\frac{1}{N}\sum_{t=1}^N \sI_\phi(X^{0}_{-t};Y_0|Y^{-1}_{-t}) \\
    &= \lim_{N\to\infty}\frac{1}{N}\sum_{t=1}^N \sI_\phi(X^{t};Y_t|Y^{t-1})\\
    &= \lim_{N\to\infty}\frac{1}{N}\sI_\phi(X^{N}\to Y^N)\\
    &= \sI_\phi(\XX\to\YY),
\end{align*}
where the third equality uses the stationarity of the joint distribution.
This concludes the proof. $\hfill\square$

\subsection{Proof of Lemma \ref{lemma:MINE_grad}}\label{proof:mine_based_grad}
Recall that we focus on the loss function:
\begin{equation}
    \hat{\sJ}^{\mathsf{MI}}_\theta(\Dn,\phi) := \frac{1}{n}\sum_{t=1}^{n}\log\left( p^\phi(X_t)\right)\left(g_\theta(X_t, Y_t) - \hat{\sI}_{\mathsf{MI}}(\Dn)\right).
\end{equation}
First consider the partial derivative w.r.t. a single coordinate.
Assume, without loss of generality, that $\cX=[1,\dots,m]$ and fix $i\in[1,\dots,m]$.
Taking the derivative w.r.t. $\phi_i$, we have
\begin{align*}
    \partial_{\phi_i}\hat{\sJ}^{\mathsf{MI}}_\theta(\Dn,\phi) = \frac{1}{n}\sum_{t=1}^{n}\partial_{\phi_i}\log\left( p^\phi(X_t)\right)\left(g_\theta(X_t, Y_t) - \hat{\sI}_{\mathsf{MI}}(\Dn)\right). 
\end{align*}
Since $p^\phi(X_t) = \sigma_{\mathsf{sm}}^{X_t}(\phi)$, we obtain
\begin{align*}
    \partial_{\phi_i}\log p^\phi(X_t) &= \partial_{\phi_i}\log e^{\phi_{X_t}}-\partial_{\phi_i}\log\left( \sum_{k=1}^{m}e^{\phi_k} \right)\\
    &= \partial_{\phi_i}\phi_{X_t} - \frac{e^{\phi_i}}{\sum_{k=1}^{m} e^{\phi_k}}\\
    &= \mathbbm{1}_{\{X_t=i\}} - \sigma_{\mathsf{sm}}^{X_t}(\phi).
\end{align*}
Therefore, the gradient is given by
\begin{equation}
    \nabla_\phi\hat{\sJ}^{\mathsf{MI}}_\theta(\Dn,\phi) = \frac{1}{n}\sum_{t=1}^{n}\left(e_{X_t} -  p^\phi\right)\left(g_\theta(X_t, Y_t) - \hat{\sI}_{\mathsf{MI}}(\Dn)\right).
\end{equation}
$\hfill\square$


\subsection{Proof of Lemma \ref{lemma:GP_complex_bound}}\label{appendix:GP_complex_bound}
To simplify notation, let $m:=|\cQ_g|$ and fix $m$.
Note that the number of graphs identical up to $y$ labeling is $(m)^{|\cY|m}$.
The authors of \cite{sabag2020graph} claim that the graph-pooling method reduces the number of $Q$-graphs the algorithm considers by a factor of $m!$.
Therefore, 
\begin{equation}\label{eq:gp_init}
    N_{\sf{GP}} = \frac{m^{m|\cY|}}{m!}.
\end{equation}
By the Stirling upper bound approximation of $m!$ we have
\begin{align*}
     N_{\sf{GP}} &\geq \frac{m^{m|\cY|}}{\sqrt{e^2m}\frac{m^{m}}{e^{m}}}\\
     & = e^{m}m^{m(|\cY|-1)-0.5}e^{-1}\\
     &\geq e^{m}m^{m-0.5}\\
     &\geq e^{m((m-0.5)\log m)}\\
     & \geq e^{m\log m}.
\end{align*}

\section{Concluding Remarks and Future Directions}\label{sec:conclusion}
This work developed an optimization method for the estimated DI rate over communication channels with discrete input alphabets.
We proposed a deep generative model for the input PMF and derived an alternative optimization objective which easy to differentiate w.r.t. the parameters of the PMF model. This new objective was derived via an MDP formulation of the DI optimization problem, combined with the policy gradients method and DINE-based function approximation.
The overall procedure is an iterative estimation-optimization routine of the DI, where the estimation step involves training the DINE. To the best of our knowledge, ours is the first method that can optimize estimated DI over discrete inputs when the channel model is unknown.


To demonstrate the utility of our approach, we used it to estimate the capacities of various channels, under both feedforward and feedback communication schemes.
The capacity estimates demonstrated significant correspondence with known theoretical solutions and/or bounds, and the learned input PMF was shown to coincide with capacity-achieving input distributions.
In addition, we showed how to leverage the optimized input PMF model to calculate lower and upper bounds on the feedback capacity of unifilar FSCs via $Q$-graphs. Lastly, we demonstrated how our algorithm gives rise to probabilistic shaping schemes of PAM and QAM constellations for the PP-AWGN.

Our work enables the optimization of estimated DI, treating the channel as a black-box that can be sampled.
This method is beneficial for data-driven time-series tasks, when control over some of its elements is assumed.
In future work, we aim to apply the proposed scheme to multi-user communication channels by generalizing our framework to multi-agent reinforcement learning. We will also explore applications to sequential machine learning and sequential control.

\newpage
\bibliographystyle{unsrt}
\bibliography{main.bib}
\end{document}